\documentstyle[aps,eqsecnum,multicol,amsfonts,pra]{revtex}

\begin{document}
\draft

\title{Phase-space formulation of quantum mechanics and quantum 
state reconstruction for physical systems with Lie-group symmetries}
\author{C. Brif \cite{email1} and A. Mann \cite{email2}}
\address{Department of Physics, Technion -- Israel Institute of 
Technology, Haifa 32000, Israel}
\maketitle

\begin{abstract}
We present a detailed discussion of a general theory of phase-space
distributions, introduced recently by the authors
[J. Phys. A {\bf 31}, L9 (1998)]. This theory provides a unified 
phase-space formulation of quantum mechanics for physical systems 
possessing Lie-group symmetries. The concept of generalized coherent 
states and the method of harmonic analysis are used to construct 
explicitly a family of phase-space functions which are postulated 
to satisfy the Stratonovich-Weyl correspondence with a generalized 
traciality condition. The symbol calculus for the phase-space 
functions is given by means of the generalized twisted product. 
The phase-space formalism is used to study the problem of the 
reconstruction of quantum states. In particular, we consider the 
reconstruction method based on measurements of displaced projectors, 
which comprises a number of recently proposed quantum-optical schemes 
and is also related to the standard methods of signal processing.
A general group-theoretic description of this method is developed
using the technique of harmonic expansions on the phase space.
\end{abstract}

\pacs{03.65.Bz, 03.65.Fd}

\begin{multicols}{2}

\section{Introduction}

The phase-space formulation of quantum mechanics has a long history. 
In 1932 Wigner \cite{Wig32} introduced his famous function which has 
found numerous applications in many areas of physics and electronics. 
In 1949 Moyal \cite{Moy49} discovered that the Weyl correspondence
rule \cite{Weyl} can be inverted by the Wigner transform from an
operator on the Hilbert space to a function on the phase space. 
As a result, the quantum expectation value of an operator can be 
represented by the statistical-like average of the corresponding 
phase-space function with the statistical density given by the Wigner 
function associated with the density matrix of the quantum state.
In this way quantum mechanics can be formally represented as a
statistical theory on classical phase space. It should be
emphasized that this phase-space formalism does not replace quantum 
mechanics by a classical or semiclassical theory. In fact, the
phase-space formulation of quantum mechanics (also known as the Moyal 
quantization) is in principle equivalent to conventional formulations
due to Heisenberg, Schr\"{o}dinger, and Feynman. However, the formal
resemblance of quantum mechanics in the Moyal formulation to
classical statistical mechanics can yield deeper understanding of
differences between the quantum and classical theories. 
Extensive lists of the literature on this subject can be found in
reviews and books \cite{CaZa83,Hill84,Kim91,Lee95,Gade95,Schroek96}.

The ideas of Moyal were further developed in the late sixties in 
the works of Cahill and Glauber \cite{CaGl69} and Agarwal and Wolf 
\cite{AgWo70}. As mentioned, the Wigner function is related to 
the Weyl (symmetric) ordering of the position and momentum 
operators $q$ and $p$ or, equivalently, of the bosonic 
annihilation and creation operators $a$ and $a^{\dagger}$. 
However, there exist other possibilities of ordering. 
In particular, it was shown \cite{CaGl69} that the 
Glauber-Sudarshan $P$ function \cite{Gla63,Sudar63} is associated 
with the normal ordering and the Husimi $Q$ function 
\cite{Husimi} with the antinormal ordering 
of $a$ and $a^{\dagger}$. Moreover, a whole family of 
$s$-parameterized functions can be defined on the complex plane 
which is equivalent to the $q$-$p$ flat phase space. The index
$s$ is related to the corresponding ordering procedure of $a$ and 
$a^{\dagger}$; the values $+1$, $0$, and $-1$ of $s$ correspond
to the $P$, $W$, and $Q$ functions, respectively. 
These phase-space functions are referred to as quasiprobability 
distributions (QPDs), as they play in the Moyal formulation of 
quantum mechanics a role similar to that of genuine probability 
distributions in classical statistical mechanics. Various QPDs 
has been extensively used in many quantum-optical 
applications \cite{Gardiner,Perina}.
Most recently, there is great interest in the $s$-parameterized 
distributions because of their role in modern schemes for 
measuring the quantum state of the radiation field 
\cite{Leon:book}.  

The mathematical framework and the conceptual background of the 
Moyal quantization have been essentially enlarged and generalized 
in two important papers by Bayen {\em et al.} \cite{Bay78}.
Specifically, it was shown that noncommutative deformations of 
the algebra of classical phase-space functions (defined by the
ordinary multiplication) give rise to operator algebras of quantum
mechanics. This fact means that introducing noncommutative symbol 
calculus based on the so-called twisted product (also known as 
the star or Moyal product), one obtains a completely autonomous 
reformulation of quantum mechanics in terms of phase-space 
functions instead of Hilbert-space states and operators. This 
program of ``quantization by deformation'' has been developed in 
a number of works \cite{Fron79a,Fron79b,Huynh,BFLS84,MoON}.

For a long time applications of the Moyal formulation were
restricted to description of systems like a spinless
non-relativistic quantum particle or a mode of the quantized
radiation field (modeled by a quantum harmonic oscillator),
i.e., to the case of the flat phase space. Therefore, an 
important problem is the generalization of the standard Moyal 
quantization for quantum systems possessing an intrinsic group of 
symmetries, with the phase space being a homogeneous manifold on 
which the group of transformations acts transitively
\cite{Fron79b,Anton98}. 
It has been recently understood that this problem can be solved 
using the Stratonovich-Weyl (SW) correspondence. The idea of the 
SW correspondence is that the linear bijective mapping between
operators on the Hilbert space and functions on the phase space
can be implemented by a kernel which satisfies a number of
physically sensible postulates, with covariance and traciality
being the two most important ones. This idea first appeared in a 
paper by Stratonovich \cite{Stra56} in 1956, but it was almost 
forgotten for decades. The SW correspondence, that has been 
restated some years ago by Gracia-Bond\'{\i}a and V\'{a}rilly
\cite{GBVa88,VaGB89}, has given a new impulse to the phase-space
formulation of quantum theory. The SW method of the Moyal 
quantization has been applied to a number of important 
situations: a nonrelativistic free particle with spin, using the 
extended Galilei group \cite{GBVa88}; a relativistic free 
particle with spin, using the Poincar\'{e} group \cite{CaGBVa90}; 
the spin, using the SU(2) group \cite{VaGB89}; compact semisimple 
Lie groups \cite{FiGBVa90}; one- and two-dimensional kinematical 
groups \cite{GMNO91,BGO92,MaOl96,ArOl97a}; the two-dimensional
Euclidean group \cite{GMNO91,ArOl97b}; systems of identical
quantum particles \cite{GaNi}. For a review of basic results see
Ref.\ \cite{Gade95}.

Notwithstanding the success of the SW method in the Moyal 
quantization of many important physical systems, the theory
suffered from a serious problem. Specifically, it was the absence 
of a simple and effective method for the construction of the SW
kernel which should implement the mapping between Hilbert-space 
operators and phase-space functions. The construction procedures
for the SW kernels, considered during the last decade (see, e.g.,
Ref.\ \cite{Gade95}), did not guarantee that the kernel will
satisfy all the SW postulates. Only very recently a general 
algorithm for constructing the SW kernel for quantum systems 
possessing Lie-group symmetries was proposed \cite{BM98}. 
It has been shown that the constructed kernel explicitly
satisfies all the desired properties (the SW postulates) and that
in the particular cases of the Heisenberg-Weyl group and SU(2) our 
general expression reduces to the known results. 

In the present paper we essentially extend the results of Ref.\ 
\cite{BM98} and present a self-consistent theory of the SW method
for the phase-space formulation of quantum mechanics. This theory
makes use of the concept of generalized coherent states and of
some basic ideas of harmonic analysis. Like the Cahill-Glauber 
formalism for the Heisenberg-Weyl group, we construct the 
$s$-parameterized family of functions on the phase space 
of a quantum system whose dynamical symmetry group is an arbitrary 
(finite-dimensional) Lie group. Accordingly, we introduce
$s$-generalized versions of the traciality condition and the
twisted product. The developed phase-space formulation is used
for a general group-theoretic description of the quantum state 
reconstruction method. This description can be useful not only for 
measurements of quantum states but also in the field of signal 
processing.

\section{Basics of Moyal quantization} 

\subsection{Generalized coherent states and the definition of 
quantum phase space}

Given a specific physical system, the first thing one needs to do for 
the Moyal quantization (i.e., for constructing phase-space functions) 
is to determine what is the related phase space. This can often 
be done by analogy with the corresponding classical problem,
thereby providing a direct route for the quantum-classical 
correspondence. From the technical point of view, the phase space can
be conveniently determined using the concept of coherent states
\cite{Per}. The coherent-state approach is not just a convenient 
mathematical tool, but it also helps to understand how physical 
properties of the system are reflected by the geometrical structure 
of the related phase space. It is possible to say that the concept 
of coherent states constitutes a bridge between the Moyal phase-space 
quantization and the Berezin geometric quantization \cite{Ber75}.

Let $G$ be a Lie group (connected and simply connected, with finite 
dimension $n$), which is the dynamical symmetry group of a given 
quantum system. Let $T$ be a unitary irreducible representation of 
$G$ acting on the Hilbert space ${\cal H}$. By choosing a fixed 
normalized reference state $|\psi_0\rangle \in {\cal H}$, one can 
define the system of coherent states $\{|\psi_g\rangle\}$:
\begin{equation}
|\psi_g\rangle = T(g) |\psi_0\rangle , \hspace{0.8cm}  g \in G .
\end{equation}
The isotropy subgroup $H \subset G$ consists of all the group 
elements $h$ that leave the reference state invariant up to a 
phase factor,
\begin{equation}
T(h) |\psi_{0}\rangle = e^{ i \phi(h)} |\psi_{0}\rangle , 
\hspace{0.8cm} | e^{ i \phi(h)} | = 1 ,
\hspace{0.3cm} h \in H .                
\end{equation}
For every element $g \in G$, there is a decomposition 
of $g$ into a product of two group elements, one in $H$ and 
the other in the coset space $X = G/H$,
\begin{equation}
g = \Omega h , \hspace{0.8cm} g \in G , \;\; h \in H , \;\; 
\Omega \in X .  
\end{equation}
It is clear that group elements $g$ and $g'$ with different $h$ 
and $h'$ but with the same $\Omega$ produce coherent states 
which differ only by a phase factor: 
$|\psi_{g}\rangle = e^{ i \delta} |\psi_{g'}\rangle$, where 
$\delta =\phi(h) -\phi(h')$. 
Therefore a coherent state $|\Omega\rangle \equiv
|\psi_{\Omega}\rangle$ is 
determined by a point $\Omega = \Omega(g)$ in the coset 
space $X$. A very important property is the identity 
resolution in terms of the coherent states:
\begin{equation}
\int_{X} d \mu(\Omega) |\Omega\rangle\langle\Omega| = I ,
\label{idres}
\end{equation}
where $d \mu(\Omega)$ is the invariant integration measure
on $X$, the integration is over the whole manifold $X$, 
and $I$ is the identity operator on ${\cal H}$. The natural
action of $G$ on $X$ will be denoted by $g\cdot\Omega$.

An important class of coherent-state systems 
corresponds to the coset spaces $X = G/H$ which are homogeneous
K\"{a}hlerian manifolds. Then $X$ can be considered as the phase 
space of a classical dynamical system, and the mapping 
$\Omega \rightarrow |\Omega \rangle\langle \Omega|$ 
is the geometric quantization for this system \cite{Ber75}. 
The standard (or maximum-symmetry) systems of the coherent states 
correspond to the cases when an `extreme' state of the 
representation Hilbert space (e.g., the vacuum state of an 
oscillator or the lowest/highest spin state) is chosen as the 
reference state. This choice of the reference state leads to 
systems consisting of states with properties ``closest to those 
of classical states'' \cite{Per,BMR98}. In what follows we will 
consider the coherent states of maximal symmetry and assume that 
the phase space of the quantum system is a homogeneous 
K\"{a}hlerian manifold $X = G/H$, each point of which corresponds 
to a coherent state $|\Omega\rangle$. In particular, the Glauber
coherent states of the Heisenberg-Weyl group H$_{3}$ are 
defined on the complex plane 
${\Bbb{C}} = {\rm H}_3 / {\rm U}(1)$, and the spin coherent
states are defined on the unit sphere 
${\Bbb{S}}^2 = {\rm SU}(2) / {\rm U}(1)$.
In the more rigorous mathematical language of Kirillov's theory
\cite{Kiril}, the phase space $X$ is defined as the coadjoint 
orbit associated with the unitary irreducible representation $T$
of the group $G$ on the Hilbert space ${\cal H}$. 

\subsection{The Stratonovich-Weyl correspondence} 

Once the phase space of a quantum system is determined, the Moyal
quantization proceeds in the following way. Let $A$ be an operator 
on ${\cal H}$. Then $A$ can be mapped by a family of functions 
$F_{A}(\Omega;s)$ onto the phase space $X$ (the index $s$ labels 
functions in the family). If $A$ is the density matrix $\rho$ of 
a quantum system, the corresponding phase-space functions
$F_{\rho}(\Omega;s) \equiv P(\Omega;s)$ are called QPDs.
Of course, the phase-space formulation of the quantum theory for 
a given physical system can be successful only if the functions 
$F_{A}(\Omega;s)$ possess some physically motivated properties. 
These properties were formulated by Stratonovich \cite{Stra56} and 
are referred to as the SW correspondence:
\begin{enumerate}
\begin{mathletters}
\item[(0)] Linearity: $A \rightarrow F_{A}(\Omega;s)$ 
is one-to-one linear map.
\item[(i)] Reality: 
\begin{equation}
F_{A^{\dagger}}(\Omega;s) = [F_{A}(\Omega;s)]^{\ast} .
\label{real} 
\end{equation}
\item[(ii)] Standardization: 
\begin{equation}
\int_{X} d \mu(\Omega) F_{A}(\Omega;s) = {\rm Tr}\, A .
\label{stand}
\end{equation}
\item[(iii)] Covariance: 
\begin{equation}
F_{A(g)}(\Omega;s) = F_{A}(g\cdot\Omega;s) , 
\label{covar}
\end{equation}
where $A(g) \equiv T(g^{-1}) A T(g)$.
\item[(iv)] Traciality:
\begin{equation}
\int_{X} d \mu(\Omega) F_{A}(\Omega;s) 
F_{B}(\Omega;-s) = {\rm Tr}\, (AB) .
\label{trac}
\end{equation}
\end{mathletters}
\end{enumerate}
If the function $F_{A}(\Omega;s)$ satisfies the SW correspondence,
it is called the SW symbol of the operator $A$.

The above conditions have a clear physical meaning. The linearity 
and the traciality conditions are related to the statistical 
interpretation of the theory. If $B$ is the density matrix (the 
state operator) of a system, then the traciality condition 
(\ref{trac}) assures that the statistical average of the phase-space 
distribution $F_{A}$ coincides with the quantum expectation value 
of the operator $A$. O'Connell and Wigner \cite{OCWi81} have shown 
that the traciality condition for density matrices of a spinless 
quantum particle (there it appears as an overlap relation) is 
necessary for the uniqueness of the definition of the Wigner 
function. 
It has been also shown \cite{FiGBVa90} that the traciality 
condition is necessary for the uniqueness of the definition of 
the symbol calculus (twisted or ``star'' products) of the 
phase-space functions and for the validity of the related 
non-commutative Fourier analysis. 
Equation (\ref{trac}) is actually a generalization of the 
usual traciality condition \cite{Stra56,VaGB89,FiGBVa90}, as it
holds for any $s$ and not only for the Wigner case $s=0$.
The reality condition (\ref{real}) means that
if $A$ is self-adjoint, then $F_{A}(\Omega;s)$ is real.
The condition (\ref{stand}) is a natural normalization,
which means that the image of the identity operator $I$ is
the constant function $1$. The covariance condition
(\ref{covar}) means that the phase-space formulation
must explicitly express the symmetry of the system.

The linearity is taken into account if we implement the map
$A \rightarrow F_{A}(\Omega;s)$ by the generalized Weyl 
rule
\begin{equation}
F_{A}(\Omega;s) = {\rm Tr}\, [A \Delta(\Omega;s)] ,
\label{gwr}
\end{equation}
where $\{\Delta(\Omega;s)\}$ is a family (labeled by $s$)
of operator-valued functions on the phase space $X$. 
These operators are referred to as the SW kernels. 
The generalized traciality condition (\ref{trac}) is taken 
into account if we define the inverse of the generalized 
Weyl rule (\ref{gwr}) as
\begin{equation}
A = \int_{X} d \mu(\Omega) F_{A}(\Omega;s) 
\Delta(\Omega;-s) . \label{invmap}
\end{equation}
Now, the conditions (\ref{real})-(\ref{covar}) of the SW 
correspondence for $F_{A}(\Omega;s)$ can be translated 
into the following conditions on the SW kernel 
$\Delta(\Omega;s)$:
\begin{mathletters}
\begin{eqnarray}
\rm{(i)} & & \;\;\; \Delta(\Omega;s) = 
[\Delta(\Omega;s)]^{\dagger}
\;\;\;\;\;\; \forall \Omega \in X .
\label{real1} \\
\rm{(ii)} & & \;\;\; \int_{X} d \mu(\Omega) \Delta(\Omega;s) 
= I . \label{stand1} \\
\rm{(iii)} & & \;\;\; \Delta(g\cdot\Omega;s) = 
T(g) \Delta(\Omega;s) T(g^{-1}) . 
\label{covar1}
\end{eqnarray}
\end{mathletters}

Substituting the inverted maps (\ref{invmap}) for $A$ and 
$B$ into the generalized traciality condition (\ref{trac}),
we obtain the relation between functions with different 
values of the index $s$: 
\begin{eqnarray}
& & F_{A}(\Omega;s) = \int_{X} d \mu(\Omega') 
K_{s,s'}(\Omega,\Omega') F_{A}(\Omega';s') , 
\label{genrel} \\
& & K_{s,s'}(\Omega,\Omega') \equiv 
{\rm Tr}\, [ \Delta(\Omega;s) \Delta(\Omega';-s')] .
\label{Kssfun}
\end{eqnarray}
If we take $s=s'$ in Eq.~(\ref{genrel}) and take into account
the arbitrariness of $A$, we obtain the following relation
\begin{equation}
\Delta(\Omega;s) = \int_{X} d \mu(\Omega') K(\Omega,\Omega')
\Delta(\Omega';s) ,  \label{specrel}
\end{equation}
where the function
\begin{equation}
K(\Omega,\Omega') = {\rm Tr}\, [ \Delta(\Omega;s) 
\Delta(\Omega';-s)]
\label{Kfun}
\end{equation}
behaves like the delta function on the manifold $X$.

\section{Construction of the Stratonovich-Weyl kernel}

It is clear that the Moyal quantization for a physical system 
is accomplished by constructing the SW kernel $\Delta(\Omega;s)$ 
that satisfies the SW postulates. Although the form of the SW 
kernel has been known for many systems, a general construction 
method was not known. A procedure that was 
applied in many works \cite{Gade95,GMNO91,BGO92,MaOl96}
is as follows. An arbitrary point 
$\Omega_{0} \in X$ is fixed and then an Ansatz is made for
a self-adjoint operator $\Delta(\Omega_{0})$ (usually only the 
case $s=0$ was considered) that satisfies the standardization 
condition (\ref{stand1}) and the following property:
\begin{equation}
\Delta(\Omega_{0}) = T(\gamma) \Delta(\Omega_{0}) T(\gamma^{-1}) ,
\hspace{0.8cm} \forall \gamma \in H_{\Omega_{0}} ,
\end{equation}
where $H_{\Omega_{0}} = \{ \gamma \in G \ | \  
\gamma\cdot\Omega_{0} = \Omega_{0} \}$ is the isotropy subgroup 
for $\Omega_{0}$. For any $\Omega \in X$ there exists $g \in G$ 
such that $g\cdot\Omega_{0} = \Omega$, and then the SW kernel is 
defined by
\begin{equation}
\Delta(\Omega) = \Delta(g\cdot\Omega_{0}) = 
T(g) \Delta(\Omega_{0}) T(g^{-1}) .
\end{equation}
This kernel automatically satisfies the covariance condition
(\ref{covar1}), but the problem is that the traciality is not
guaranteed. Of course, in the described procedure the form of 
the kernel depends on the Ansatz and often no kernel 
satisfying the traciality condition is found.

We propose here a simple and general algorithm for constructing 
the SW kernels (the whole $s$-parameterized family) which 
explicitly satisfy all the SW postulates, including both the
covariance and the traciality. Our method makes use of 
Perelomov's concept of coherent states and of only some 
basic ideas from harmonic analysis. Hopefully, the simplicity
and generality of our method will draw more attention to the 
ideas of the phase-space quantization.

\subsection{Necessary instruments: harmonic functions,
invariant coefficients, and tensor operators}

Our problem is to find the explicit form of the SW kernel 
$\Delta(\Omega;s)$ that satisfies the conditions 
(\ref{real1})-(\ref{covar1}) and (\ref{specrel}). In order
to accomplish this task, we need three basic ingredients:
harmonic functions, invariant coefficients, and tensor 
operators. The coherent states serve here as the glue
that binds them together.

We start by considering the Hilbert space $L^{2}(X,\mu)$ of 
square-integrable functions $u(\Omega)$ on $X$ with the 
invariant measure $ d \mu$. The representation $T$ of the 
Lie group $G$ on $L^{2}(X,\mu)$ is defined as 
\begin{equation}
T(g) u(\Omega) = u(g^{-1}\cdot\Omega) . 
\end{equation}
The eigenfunctions $Y_{\nu}(\Omega)$ of the Laplace-Beltrami
operator \cite{BaRa86} form a complete orthonormal basis in
$L^{2}(X,\mu)$:
\begin{mathletters}
\begin{eqnarray}
& & \sum_{\nu} Y_{\nu}^{\ast}(\Omega) Y_{\nu}(\Omega') 
= \delta(\Omega-\Omega') , \label{completeness} \\
& & \int_{X} d \mu(\Omega) Y_{\nu}^{\ast}(\Omega) 
Y_{\nu'}(\Omega) = \delta_{\nu \nu'} . \label{orthonormality}
\end{eqnarray}
\end{mathletters}
The functions $Y_{\nu}(\Omega)$ are called the harmonic 
functions, and $\delta(\Omega-\Omega')$ is the delta function in 
$X$ with respect to the measure $ d \mu$. 
Note that the index $\nu$ is multiple; it has one discrete part, 
while the other part is discrete for compact manifolds and 
continuous for noncompact manifolds. In the latter case the 
summation over $\nu$ includes an integration with the Plancherel 
measure $ d \rho(\nu)$ and the symbol $\delta_{\nu \nu'}$ 
includes some Dirac delta functions (for more details see Ref.\ 
\cite{BaRa86}). For conciseness, we omit these details in our 
formulas. 
The harmonic functions $Y_{\nu}(\Omega)$ are linear
combinations of matrix elements $T_{\nu \nu'}(g)$. 
Therefore, the transformation rule for the harmonic 
functions is \cite{BaRa86}
\begin{equation}
T(g) Y_{\nu}(\Omega) = Y_{\nu}(g^{-1}\cdot\Omega)
= \sum_{\nu'} T_{\nu' \nu}(g) Y_{\nu'}(\Omega) .
\label{transrule}
\end{equation}
It should be understood that the summation in 
Eq.~(\ref{transrule}) is only on the part of $\nu$ that labels
functions within an irreducible subspace.

Next, we once again use the coherent states, in order to 
introduce the concept of invariant coefficients. 
The positive-valued function 
$|\langle \Omega | \Omega' \rangle|^2$ is symmetric in $\Omega$ 
and $\Omega'$. Therefore, its expansion in the orthonormal basis 
must be of the form
\begin{eqnarray}
|\langle \Omega | \Omega' \rangle|^2 & = & \sum_{\nu} \tau_{\nu} 
Y_{\nu}^{\ast}(\Omega) Y_{\nu}(\Omega') \nonumber \\
& = & \sum_{\nu} \tau_{\nu} Y_{\nu}^{\ast}(\Omega') 
Y_{\nu}(\Omega) ,
\label{ooexpansion}
\end{eqnarray}
where $\tau_{\nu}$ are real positive coefficients.
Using the invariance $\langle \Omega | \Omega' \rangle = 
\langle g\cdot\Omega | g\cdot\Omega' \rangle$ and the unitarity 
of the representation $T$, we obtain
\begin{eqnarray}
|\langle \Omega | \Omega' \rangle|^2 & = & \sum_{\nu} \tau_{\nu} 
Y_{\nu}^{\ast}(g\cdot\Omega) Y_{\nu}(g\cdot\Omega') \nonumber \\
& = & \sum_{\nu'} Y_{\nu'}^{\ast}(\Omega) \sum_{\nu} \tau_{\nu} 
T_{\nu \nu'}(g) Y_{\nu}(g\cdot\Omega') .
\end{eqnarray}
In order to satisfy this equality, the coefficients 
$\tau_{\nu}$ must be invariant under the index 
transformation of equation (\ref{transrule}): 
$\tau_{\nu} = \tau_{\nu'}$. This means that $\tau_{\nu}$ do
not depend on the part of $\nu$ which labels functions within 
an irreducible subspace.
Since the Laplace-Beltrami operator is self-adjoint, one finds
that 
\begin{equation}
Y_{\nu}^{\ast}(\Omega) = e^{ i \phi(\nu)} 
Y_{\tilde{\nu}}(\Omega) , 
\label{conjrule}
\end{equation}
where $Y_{\tilde{\nu}}(\Omega)$ is another harmonic function,
with the same eigenvalue as $Y_{\nu}(\Omega)$. 
Since $|\langle \Omega | \Omega' \rangle|^2$ is real, the
coefficients $\tau_{\nu}$ must be invariant under the
index transformation of equation (\ref{conjrule}): 
$\tau_{\nu} = \tau_{\tilde{\nu}}$.

Next we use the coherent states, harmonic functions, and 
invariant coefficients for defining the set of operators 
$\{ D_{\nu} \}$ on ${\cal H}$:
\begin{equation}
D_{\nu} \equiv \tau_{\nu}^{-1/2} \int_{X} d \mu(\Omega) 
Y_{\nu}(\Omega) | \Omega \rangle \langle \Omega | .
\label{Ddef}
\end{equation}
Using the expression (\ref{ooexpansion}) and the orthonormality 
relation (\ref{orthonormality}) for the harmonic functions, we 
obtain the orthonormality condition for the operators $D_{\nu}$:
\begin{equation}
{\rm Tr}\, (D_{\nu} D_{\nu'}^{\dagger}) = \delta_{\nu \nu'} .
\end{equation}
Note that the factor $\tau_{\nu}^{-1/2}$ in front 
of the integral in Eq.~(\ref{Ddef}) serves just for the proper
normalization. 
Using (\ref{ooexpansion}), we also obtain the relation
\begin{equation}
\tau_{\nu}^{-1/2} \langle \Omega | D_{\nu} | \Omega \rangle =
Y_{\nu}(\Omega) .
\end{equation}
The invariance of the coefficients $\tau_{\nu}$ implies that 
$D_{\nu}$ are the tensor operators whose transformation rule is 
the same as for the harmonic functions $Y_{\nu}(\Omega)$:
\begin{equation}
T(g) D_{\nu} T(g^{-1}) = 
\sum_{\nu'} T_{\nu' \nu}(g) D_{\nu'} .
\label{dtrule}
\end{equation}
A useful property of the tensor operators is that any operator 
$A$ on ${\cal H}$ can be expanded in the orthonormal basis 
$\{ D_{\nu} \}$:
\begin{equation}
A = \sum_{\nu} {\rm Tr}\, (A D_{\nu}^{\dagger}) D_{\nu} .
\label{Dexpansion}
\end{equation}

\subsection{Explicit form of the kernel}

Using the above preliminary results, we are able to find the 
SW kernel $\Delta(\Omega;s)$ with all the desired 
properties. Specifically, let us define
\begin{equation}
\Delta(\Omega;s) \equiv \sum_{\nu} f(s;\tau_{\nu})
Y_{\nu}^{\ast}(\Omega) D_{\nu} .
\label{kerneldef}
\end{equation}
We will show that the construction of the generalized kernel 
(\ref{kerneldef}) satisfies the SW correspondence.   
In equation (\ref{kerneldef}) $f(s;\tau_{\nu})$ is a function 
of $\tau_{\nu}$ and of the index $s$. 
We assume that $f$ possesses the invariance properties of 
$\tau_{\nu}$. 

Using the invariance of $\tau_{\nu}$ under the index 
transformation of Eq.\ (\ref{conjrule}), we see that the reality 
condition (\ref{real1}) is satisfied if $f(s;\tau_{\nu})$ is a 
real-valued function. Therefore, we can consider only real values 
of the index $s$. 

Next we consider the standardization condition (\ref{stand1}).
Using the definition (\ref{kerneldef}), we obtain
\begin{equation}
\int_{X} d \mu(\Omega) \Delta(\Omega;s) = 
\sum_{\nu} f(s;\tau_{\nu}) D_{\nu} \int_{X} d \mu(\Omega)
Y_{\nu}^{\ast}(\Omega) ,
\label{lside1}
\end{equation}
while Eq.\ (\ref{Dexpansion}) can be used to write
\begin{equation}
I = \sum_{\nu} {\rm Tr}\, (D_{\nu}^{\dagger}) D_{\nu}
= \sum_{\nu} \tau_{\nu}^{-1/2} D_{\nu} \int_{X} d \mu(\Omega)
Y_{\nu}^{\ast}(\Omega) .
\label{rside1}
\end{equation}
The standardization condition is satisfied if the expressions
(\ref{lside1}) and (\ref{rside1}) are equal.
Using the identity resolution (\ref{idres}) and Eq.\
(\ref{ooexpansion}), we can write
\begin{eqnarray}
1 & = & \langle \Omega | \Omega \rangle = \int_{X} d \mu(\Omega')
|\langle \Omega | \Omega' \rangle|^{2} \nonumber \\
& = & \sum_{\nu} \tau_{\nu} Y_{\nu}^{\ast}(\Omega) 
\int_{X} d \mu(\Omega') Y_{\nu}(\Omega') .
\end{eqnarray}
Multiplying the left and right sides of this equation by 
$Y_{\nu'}(\Omega)$ and integrating over $ d \mu(\Omega)$, 
we obtain
\begin{equation}
\int_{X} d \mu(\Omega) Y_{\nu}(\Omega) =
\tau_{\nu} \int_{X} d \mu(\Omega) Y_{\nu}(\Omega) .
\end{equation}
Since $\tau_{\nu}$ is not identically $1$, this relation can
be satisfied only if there exists some $\nu_{0}$ such that
$\tau_{\nu_{0}} = 1$ and
\begin{equation}
\int_{X} d \mu(\Omega) Y_{\nu}(\Omega) \propto 
\delta_{\nu \nu_{0}} .
\label{delta0}
\end{equation}
(As was already mentioned, for noncompact manifolds the symbol 
$\delta_{\nu \nu'}$ actually includes some Dirac delta functions.)
It can be easily seen from Eqs.\ (\ref{lside1}), (\ref{rside1}), 
and (\ref{delta0}) that the standardization condition 
is satisfied if $f(s;\tau_{\nu_{0}}) = \tau_{\nu_{0}}^{-1/2}$,
i.e.,
\begin{equation}
f(s;1) = 1 , \;\;\;\; \forall s .
\label{stand2}
\end{equation}

The covariance condition (\ref{covar1}) can be rewritten as
\begin{eqnarray}
& & \sum_{\nu} f(s;\tau_{\nu}) D_{\nu} Y_{\nu}^{\ast}(g\cdot\Omega)
\nonumber \\
& & = \sum_{\nu} f(s;\tau_{\nu}) T(g) D_{\nu} T(g^{-1}) 
Y_{\nu}^{\ast}(\Omega) .
\label{cov1form}
\end{eqnarray}
Using the transformation rules (\ref{transrule}) and 
(\ref{dtrule}), Eq.\ (\ref{cov1form}) can be transformed into
\begin{eqnarray}
& & \sum_{\nu} \sum_{\nu'} f(s;\tau_{\nu}) D_{\nu} 
T_{\nu \nu'}(g) Y_{\nu'}^{\ast}(\Omega) \nonumber \\ 
& & = \sum_{\nu} \sum_{\nu'} f(s;\tau_{\nu}) T_{\nu' \nu}(g) 
D_{\nu'}  Y_{\nu}^{\ast}(\Omega) .
\label{cov2form}
\end{eqnarray}
Changing the summation indexes $\nu \leftrightarrow \nu'$ on
either side of Eq.\ (\ref{cov2form}), we immediately see that
the covariance condition is satisfied by the virtue of the 
invariance of $\tau_{\nu}$ under the index transformation of
Eqs.\ (\ref{transrule}) and (\ref{dtrule}).

In order to satisfy the relation (\ref{specrel}), the
function $K(\Omega,\Omega')$ of equation (\ref{Kfun}) must
be the delta function in $X$ with respect to the measure
$ d \mu$,
\begin{equation}
K(\Omega,\Omega') = 
\sum_{\nu} Y_{\nu}^{\ast}(\Omega) Y_{\nu}(\Omega') 
= \delta(\Omega-\Omega') .
\end{equation}
This result is valid if
\begin{equation}
f(s;\tau_{\nu}) f(-s;\tau_{\nu}) = 1 .
\end{equation}
This property is satisfied only by the exponential
function of $s$, i.e.,
\begin{equation}
f(s;\tau_{\nu}) = [f(\tau_{\nu})]^{s} . \label{explaw}
\end{equation}
Note that the standardization condition (\ref{stand2})
then reads $f(1) = 1$. 

The exact form of the function $f(\tau_{\nu})$ can be determined 
if we define \cite{signofs} for $s=-1$ 
\begin{equation}
\Delta(\Omega;-1) \equiv | \Omega \rangle \langle \Omega | .
\label{Delm1}
\end{equation}
Then we obtain 
\begin{equation}
| \Omega \rangle \langle \Omega | = \sum_{\nu} 
[f(\tau_{\nu})]^{-1} Y_{\nu}^{\ast}(\Omega) D_{\nu} .
\label{Delm1a}
\end{equation}
Multiplying both sides of this equation by $Y_{\nu'}(\Omega)$
and integrating over $ d \mu(\Omega)$, we find
$f(\tau_{\nu}) = \tau_{\nu}^{-1/2}$, i.e.,
\begin{equation}
f(s;\tau_{\nu}) = \tau_{\nu}^{-s/2} .
\end{equation}
Obviously, the standardization condition
$f(1)=1$ is satisfied.
Finally, we obtain 
\begin{eqnarray}
\Delta(\Omega;s) & = & \sum_{\nu} \tau_{\nu}^{-s/2}
Y_{\nu}^{\ast}(\Omega) D_{\nu} \nonumber \\
& = & \sum_{\nu} \tau_{\nu}^{-s/2}
Y_{\nu}(\Omega) D_{\nu}^{\dagger} .
\label{kerneldeff}
\end{eqnarray}
It is evident that this kernel is completely 
determined by the harmonic functions on the corresponding 
manifold and by the coherent states which form this manifold. 
We will see that the SW kernel (\ref{kerneldeff}) is a 
generalization of the Cahill-Glauber kernel for a harmonic 
oscillator \cite{CaGl69,AgWo70} and of the Agarwal kernel for 
spin \cite{Agar81}. 

\section{Phase-space functions and the symbol calculus}

\subsection{Types of phase-space function}

As the explicit form of the SW kernels is known,
we can write the SW symbols on the phase space as
\begin{eqnarray}
F_{A}(\Omega;s) & = & \sum_{\nu} \tau_{\nu}^{-s/2} 
{\cal A}_{\nu} Y_{\nu}(\Omega) \nonumber \\
& = & \sum_{\nu} \tau_{\nu}^{-s/2} 
\tilde{\cal A}_{\nu} Y_{\nu}^{\ast}(\Omega) ,  
\label{ffform}
\end{eqnarray}
where we have defined
\begin{equation}
{\cal A}_{\nu} \equiv {\rm Tr}\, (A D_{\nu}^{\dagger}) ,
\hspace{0.8cm}
\tilde{\cal A}_{\nu} \equiv {\rm Tr}\, (A D_{\nu}) . 
\end{equation}
For a self-adjoint operator $A$, we get $\tilde{\cal A}_{\nu} =
{\cal A}_{\nu}^{\ast}$. It can be easily verified that 
substituting expressions (\ref{ffform}) and (\ref{kerneldeff}) 
into the inverse Weyl rule (\ref{invmap}), one indeed obtains
$A = \sum_{\nu} {\cal A}_{\nu} D_{\nu}$.
We also note that the function 
$K_{s,s'}(\Omega,\Omega')$ of equation (\ref{Kssfun}) is
given by
\begin{equation}
K_{s,s'}(\Omega,\Omega') = \sum_{\nu} \tau_{\nu}^{-(s-s')/2}\,
Y_{\nu}(\Omega) Y_{\nu}^{\ast}(\Omega') ,
\end{equation}
and it clearly satisfies Eq.\ (\ref{genrel}) which connects 
the functions with different values of the index $s$.
In general, let $F(\Omega)$ and $H(\Omega)$ be two 
phase-space functions such that
\begin{eqnarray}
& & F(\Omega) = \sum_{\nu} F_{\nu} Y_{\nu}(\Omega) , \\
& & H(\Omega) = \sum_{\nu} H_{\nu} Y_{\nu}(\Omega) .
\end{eqnarray}
Then they are related through the transformation
\begin{eqnarray}
& & F(\Omega) = \int_{X} d \mu(\Omega') K_{F H}(\Omega,\Omega') 
H(\Omega') , \label{FHtrans} \\
& & K_{F H}(\Omega,\Omega') = \sum_{\nu} \frac{F_{\nu}}{H_{\nu}}
Y_{\nu}(\Omega) Y_{\nu}^{\ast}(\Omega') .
\label{KFH}
\end{eqnarray}

Let $\{ |\phi_{n}\rangle \}$ be a complete orthonormal basis 
in the Hilbert space ${\cal H}$. Using the generalized Weyl 
rule (\ref{gwr}) for the operator 
$A = |\phi_{n}\rangle \langle\phi_{m}|$, we obtain
\begin{equation}
F_{A}(\Omega;s) = 
\langle\phi_{m}| \Delta(\Omega;s) |\phi_{n}\rangle
\equiv \Delta_{mn}(\Omega;s) .
\end{equation}
Using Eq.\ (\ref{kerneldeff}), we find
\begin{equation}
\Delta_{mn}(\Omega;s) = \sum_{\nu} \tau_{\nu}^{-s/2}
\langle\phi_{m}| D_{\nu}^{\dagger} |\phi_{n}\rangle 
Y_{\nu}(\Omega) .
\end{equation}
The standardization and traciality conditions (\ref{stand})
and (\ref{trac}) can be used to show that
\begin{eqnarray}
& & \int_{X} d \mu(\Omega) \Delta_{mn}(\Omega;s) = 
\delta_{mn} , \\
& & \int_{X} d \mu(\Omega) \Delta_{mn}(\Omega;s) 
\Delta_{kl}(\Omega;-s) = \delta_{ml} \delta_{nk} .
\end{eqnarray}
The functions $\Delta_{mn}(\Omega;s)$ form a useful orthonormal
basis in $L^{2}(X,\mu)$.

The SW symbols obtained for some special values of $s$ are 
frequently used in numerous applications.
In particular, for $s=-1$, we obtain the $Q$ function
(Berezin's covariant symbol \cite{Ber75}):
\begin{equation}
Q_{A}(\Omega) \equiv F_{A}(\Omega;-1) = 
\langle \Omega |A| \Omega \rangle .
\label{Qdef}
\end{equation}
Equation (\ref{Qdef}) can be easily obtained by recalling 
[see Eqs.\ (\ref{Delm1}) and (\ref{Delm1a})] that
\begin{equation}
\Delta(\Omega;-1) = | \Omega \rangle \langle \Omega |  
= \sum_{\nu} \tau_{\nu}^{1/2} Y_{\nu}^{\ast}(\Omega) D_{\nu} .
\end{equation}
For $s=1$, we obtain the $P$ function
(Berezin's contravariant symbol \cite{Ber75}):
\begin{equation}
P_{A}(\Omega) \equiv F_{A}(\Omega;1) = 
\sum_{\nu} \tau_{\nu}^{-1/2} {\cal A}_{\nu} Y_{\nu}(\Omega) , 
\end{equation}
whose defining property is
\begin{equation}
A = \int_{X} d \mu(\Omega) P_{A}(\Omega) 
| \Omega \rangle \langle \Omega | .
\end{equation}
The functions $P$ and $Q$ are counterparts in the
traciality condition (\ref{trac}).
Perhaps the most important SW symbol corresponds to $s=0$,
because this function is ``self-conjugate'' in the sense
that it is the counterpart of itself in the traciality
condition (\ref{trac}). It is natural to call the
function with $s=0$ the generalized Wigner function:
\begin{equation}
W_{A}(\Omega) \equiv F_{A}(\Omega;0) = 
\sum_{\nu} {\cal A}_{\nu} Y_{\nu}(\Omega) .
\end{equation}
The corresponding SW kernel is
\begin{equation}
\Delta(\Omega;0) \equiv \Delta_{W}(\Omega) 
= \sum_{\nu} Y_{\nu}^{\ast}(\Omega) D_{\nu} .
\end{equation}

\subsection{The generalized twisted product}

The phase-space formulation of quantum mechanics can be made
completely autonomous if one introduces a symbol calculus for
the functions on the phase space, which replaces the usual 
manipulations with operators on the Hilbert space. This symbol 
calculus is based on the so-called twisted product (or Moyal
product) which corresponds to the usual product of operators
\cite{Bay78,GBVa88,VaGB89}. 

Let us first consider the case of the Wigner function ($s=0$). 
The twisted product of two functions is denoted by 
$W_{A} \ast W_{B}$ and is determined by the condition 
\begin{equation}
W_{A}(\Omega) \ast W_{B}(\Omega) = W_{A B}(\Omega)
\label{TPdef} 
\end{equation}
for any two operators $A$ and $B$. Note that the condition 
(\ref{TPdef}) assures the associativity of the twisted product.
On the other hand, this product is, in general, noncommutative.
In this way the algebra of operators is mapped onto the algebra
of phase-space functions. If one starts from a classical
phase-space description, the introduction of the twisted product 
can be viewed as the quantization realized by a deformation of 
the algebra of functions \cite{Bay78}. 

Using the Weyl rule (\ref{gwr}) and its inverse (\ref{invmap}),
we obtain
\begin{eqnarray}
W_{A B}(\Omega) & = & {\rm Tr}\, [ \Delta_{W}(\Omega) A B ] 
\nonumber \\
& = & {\rm Tr}\, \left[ \Delta_{W}(\Omega) \int_{X} 
 d \mu(\Omega') W(\Omega') \Delta_{W}(\Omega') \right. 
\nonumber \\
& & \left. \times \int_{X} d \mu(\Omega'') W(\Omega') 
\Delta_{W}(\Omega'') \right] .
\end{eqnarray}
Introducing the function (trikernel) 
\begin{equation}
L(\Omega,\Omega',\Omega'') = {\rm Tr}\, [ 
\Delta_{W}(\Omega) \Delta_{W}(\Omega') \Delta_{W}(\Omega'') ] ,
\end{equation}
we obtain the following definition of the twisted product:
\begin{eqnarray}
(W_{A} \ast W_{B})(\Omega) & \equiv &
\int_{X} \int_{X} d \mu(\Omega') d \mu(\Omega'') 
L(\Omega,\Omega',\Omega'') \nonumber \\
& & \times W_{A}(\Omega') W_{B}(\Omega'') .
\end{eqnarray}
The so-called Moyal bracket is defined as
\begin{equation}
[W_{A},W_{B}]_{M} = - i ( W_{A} \ast W_{B} - 
W_{B} \ast W_{A} ) .
\end{equation}

The twisted product can be easily generalized for arbitrary values 
of $s$. The $s$-parameterized twisted product $(F_{A} \ast F_{B})
(\Omega;s)$ of any two functions $F_{A}(\Omega;s')$ and
$F_{B}(\Omega;s'')$ is once again determined by the condition
\begin{equation}
F_{A}(\Omega;s') \ast F_{B}(\Omega;s'') = F_{A B}(\Omega;s) .
\label{gTPdef} 
\end{equation}
Analogously to the Wigner function case, this leads to the 
definition
\begin{eqnarray}
(F_{A} \ast F_{B}) (\Omega;s) & \equiv &
\int_{X}\! \int_{X}\!\! d \mu(\Omega') d \mu(\Omega'') 
L_{s,s',s''}(\Omega,\Omega',\Omega'') \nonumber \\
& & \times F_{A}(\Omega';s') F_{B}(\Omega'';s'') ,
\end{eqnarray}
where the generalized trikernel is given by
\begin{eqnarray}
& & L_{s,s',s''}(\Omega,\Omega',\Omega'') =  {\rm Tr}\, 
[ \Delta(\Omega;s) \Delta(\Omega';-s') \Delta(\Omega'';-s'') ] 
\nonumber \\
& & = \sum_{m,n,k} \Delta_{mn}(\Omega;s) 
\Delta_{nk}(\Omega';-s') \Delta_{km}(\Omega'';-s'') .
\label{Ldef}
\end{eqnarray}

Using the standardization condition (\ref{stand1}) and the 
definition (\ref{Kssfun}), we obtain
\begin{eqnarray}
\int_{X} d \mu(\Omega) L_{s,s',s''}(\Omega,\Omega',\Omega'') 
& = & {\rm Tr}\, [ \Delta(\Omega';-s') \Delta(\Omega'';-s'') ]
\nonumber \\
& = & K_{-s',s''}(\Omega',\Omega'') .
\end{eqnarray}
This result together with the relation (\ref{genrel}) can be
used to obtain the so-called tracial identity for the 
generalized twisted product,
\begin{eqnarray}
& & \int_{X} d \mu(\Omega) (F_{A} \ast F_{B}) (\Omega;s)
\nonumber \\  & & = 
\int_{X} d \mu(\Omega) F_{A}(\Omega;s') F_{B}(\Omega;-s') ,
\label{tracid}
\end{eqnarray}
which holds for any $s$ and $s'$. Equation (\ref{tracid}) is
the phase-space version of the tracial identity for the
operators,
\begin{equation}
{\rm Tr}\, (A B) = \sum_{\nu} A_{\nu} \tilde{B}_{\nu} .
\end{equation}

Using the covariance condition (\ref{covar1}) and the 
definition (\ref{Ldef}), we find the invariance property
of the trikernel
\begin{equation}
L_{s,s',s''}(g\cdot\Omega,g\cdot\Omega',g\cdot\Omega'') 
= L_{s,s',s''}(\Omega,\Omega',\Omega'') .
\end{equation}
This property implies the equivariance of the twisted
product:
\begin{equation}
(F_{A} \ast F_{B})^{g} (\Omega;s) = 
F_{A}^{g}(\Omega;s') \ast F_{B}^{g}(\Omega;s'') ,
\end{equation}
where
\begin{equation}
F_{A}^{g} (\Omega;s) \equiv F_{A}(g^{-1}\cdot\Omega;s) .
\end{equation}

\section{Examples}

The general formalism presented above can be understood much 
better by illustrating it with a number of simple examples. We 
will consider two simple physical systems: a (nonrelativistic) 
spinless quantum particle and spin, whose dynamical symmetry 
groups are the Heisenberg-Weyl group H$_{3}$ and SU(2), 
respectively. 
It should be emphasized that the SW kernels for these basic
systems have been known for a long time \cite{note1}, so the 
novelty here is not the result itself but the method of 
derivation. Our aim is to demonstrate how the general algorithm 
works by applying it to a number of relatively simple and 
well-known problems. We will show that by identifying harmonic 
functions, invariant coefficients, and tensor operators for a 
given system, one can readily derive the explicit form of the 
SW kernel. 

\subsection{The Heisenberg-Weyl group}
\label{HWsection}

First, we consider the Heisenberg-Weyl group H$_{3}$ which is the 
dynamical symmetry group for a spinless quantum particle and for 
a mode of the quantized radiation field (modeled by a quantum 
harmonic oscillator). The Wigner function \cite{Wig32} and the 
Moyal quantization \cite{Moy49} were originally introduced for 
such systems. The kernel implementing the mapping between  
Hilbert-space operators and $s$-parameterized families of 
phase-space functions (the SW kernel in our notation) for H$_{3}$
was introduced by Cahill and Glauber \cite{CaGl69}.
The generalization of the formalism to the many-dimensional case 
is straightforward (see, e.g., Ref.\ \cite{Gade95}).

The nilpotent Lie algebra of H$_{3}$ is spanned 
by the basis $\{a,a^{\dagger},I\}$, where $a$ and 
$a^{\dagger}$ are the boson annihilation and creation
operators, satisfying the canonical commutation relation,
$[a,a^{\dagger}] = I$. Group elements can be parameterized
in the following way:
\begin{equation}
g=g(\gamma,\varphi) , \hspace{0.8cm}
T(g) = e^{\gamma a^{\dagger} - \gamma^{\ast} a} e^{ i \varphi I}  ,
\label{gparam}
\end{equation}
where $\gamma \in {\Bbb{C}}$ and $\varphi \in \Bbb{R}$.

The phase space is the complex plane 
${\Bbb{C}} = {\rm H}_3 / {\rm U}(1)$, and the (Glauber)
coherent states are 
\begin{equation}
|\Omega\rangle \equiv |\alpha \rangle =
D(\alpha) |0\rangle, 
\hspace{0.5cm} \alpha \in {\Bbb{C}}, 
\end{equation}
where 
\begin{equation}
D(\alpha) = \exp(\alpha a^{\dagger} - \alpha^{\ast} a)
\end{equation}
is the displacement operator. The invariant measure is 
\begin{equation}
 d \mu(\Omega) \equiv \pi^{-1}  d^{2} \alpha , 
\end{equation}
and the 
corresponding delta function is 
\begin{equation}
\delta(\Omega - \Omega')
\equiv \pi \delta^{(2)}(\alpha-\alpha') .
\end{equation}
The harmonic functions on ${\Bbb{C}}$ are the exponentials: 
\begin{equation}
Y_{\nu}(\Omega) \equiv Y_{\xi}(\alpha) \equiv Y(\xi,\alpha)
= \exp(\xi \alpha^{\ast} - \xi^{\ast} \alpha) . 
\label{YfunHW}
\end{equation}
Here $\nu \equiv \xi \in {\Bbb{C}}$ 
with the Plancherel measure given by
$ d \rho(\nu) \equiv \pi^{-1}  d^{2} \xi$ and with
$\delta_{\nu,\nu'} \equiv \pi \delta^{(2)}(\xi-\xi')$.
Note that for the Heisenberg-Weyl group both the 
phase-space coordinate $\Omega \equiv \alpha$ 
and the index $\nu \equiv \xi$ are complex numbers,
and the Plancherel measure is similar to the invariant 
measure on ${\Bbb{C}}$. 

The invariant coefficients $\tau_{\nu} \equiv \tau(\xi)$ 
can be found in the following way. In the present context
Eq.\ (\ref{ooexpansion}) takes the form
\begin{displaymath}
|\langle\alpha|\alpha'\rangle|^{2} = 
e^{-|\alpha - \alpha'|^{2}}
= \int_{{\Bbb{C}}} \frac{ d^{2} \xi}{\pi}  \tau(\xi)
e^{\xi^{\ast} (\alpha - \alpha') - 
\xi (\alpha -  \alpha')^{\ast}} .
\end{displaymath}
Taking into account that the Fourier transform of a 
Gaussian function is once again a Gaussian, it is not
difficult to obtain 
\begin{equation}
\tau(\xi) = \exp(-|\xi|^{2}) .
\end{equation}
Then we deduce that the tensor operator
\begin{equation}
D_{\nu} \equiv D(\xi) = e^{|\xi|^{2}/2} \int_{{\Bbb{C}}}
\frac{ d^{2} \alpha}{\pi} 
e^{\xi \alpha^{\ast} - \xi^{\ast} \alpha}
|\alpha \rangle \langle \alpha |  
\end{equation}
is just the displacement operator 
$D(\xi) = e^{\xi a^{\dagger} - \xi^{\ast} a}$.
The natural orthonormal basis in the Hilbert space 
is the Fock basis $\{ |n\rangle \}$,
$a^{\dagger} a |n\rangle = n |n\rangle$ ($n=0,1,2,\ldots$).
The matrix elements of the tensor operator are given by
\cite{Per}
\begin{eqnarray*}
& & \langle m| D(\xi) |n \rangle  \\
& & = \left\{ \begin{array}{l}
\sqrt{n!/m!}\, e^{-|\xi|^{2}/2} \xi^{m-n} 
L_{n}^{m-n}(|\xi|^{2}) , \;\;\; m \geq n \vspace*{2mm} \\ 
\sqrt{m!/n!}\, e^{-|\xi|^{2}/2} (-\xi^{\ast})^{n-m} 
L_{m}^{n-m}(|\xi|^{2}) , \;\;\; m \leq n ,
\end{array} \right. 
\end{eqnarray*}
where $L_{n}^{p}(x)$ are the associated Laguerre polynomials.
Using the parameterization (\ref{gparam}) of group elements,
one can easily find the transformation rule
\begin{eqnarray}
T(g) D(\xi) T(g^{-1}) & = & D(\gamma) D(\xi) D(-\gamma) 
\nonumber \\
& = & \exp(\gamma\xi^{\ast} - \gamma^{\ast}\xi) D(\xi) .
\label{DDDidentity}
\end{eqnarray}
Therefore, the index $\xi$ does not change under the group 
transformation, as $D(\xi)$ and $Y(\xi,\alpha)$ are just
multiplied by a phase factor. Correspondingly, there is no
index transformation, induced by the action of group elements, 
to which $\tau(\xi)$ should be invariant. On the other hand,
$Y^{\ast}(\xi,\alpha) = Y(-\xi,\alpha)$, and $\tau(\xi)$ is
obviously invariant under the index transformation 
$\xi \leftrightarrow -\xi$.

Finally, the harmonic functions $Y(\xi,\alpha)$, the invariant 
coefficients $\tau(\xi)$, and the tensor operators $D(\xi)$ 
can be substituted into the general formula (\ref{kerneldeff}). 
Then one obtains the SW kernel for the Heisenberg-Weyl group:
\begin{equation}
\Delta(\alpha;s) = \int_{{\Bbb{C}}}
\frac{ d^{2} \xi}{\pi} e^{s |\xi|^{2}/2} 
e^{\xi^{\ast} \alpha - \xi \alpha^{\ast}}
e^{\xi a^{\dagger} - \xi^{\ast} a} ,
\end{equation}
which is exactly the kernel introduced by Cahill and Glauber 
\cite{CaGl69}.

\subsection{The SU(2) group}

As another example, we consider SU(2) which is the dynamical
symmetry group for the angular momentum or spin and for many other
systems, for example, a collection of two-level atoms, the Stokes
operators describing the polarization of the quantized light field,
two light modes with a fixed total photon number, etc. 
A number of authors have used different approaches to the 
construction of the Wigner function for spin 
\cite{Fron79b,MoON,VaGB89,Ber75,Agar81,KaSu69,Gilm,DAS94,KBWolf,%
CziBen,Woot87,Leonh}. The explicit expressions for the $Q$, $W$, 
and $P$ functions for arbitrary spin were first obtained by Agarwal 
\cite{Agar81}, who used the spin coherent-state representation 
\cite{Per,Klaud,ACGT72} and the Fano multipole operators \cite{Fano}. 
V\'{a}rilly and Gracia-Bond\'{\i}a \cite{VaGB89} have shown that 
the spin coherent-state approach is equivalent to the formalism 
based on the SW correspondence.

The simple Lie algebra of SU(2) is spanned by the basis 
$\{J_{x},J_{y},J_{z}\}$, 
\begin{equation}
[J_{p},J_{r}] = i \epsilon_{p r t} J_{t} .
\end{equation}
The unitary irreducible representations are labeled by the 
index $j$ ($j=0,1/2,1,\ldots$), and the Hilbert space 
${\cal H}_{j}$ is spanned by the orthonormal basis 
$|j,\mu\rangle$ ($\mu=j,j-1,\ldots,-j$). Group elements can be 
parameterized using the Euler angles $\alpha,\beta,\gamma$:
\begin{equation}
g=g(\alpha,\beta,\gamma) = e^{ i \alpha J_{z} }  
e^{ i \beta J_{y} } e^{ i \gamma J_{z} } .
\label{gEuler}
\end{equation}

The phase space is the unit sphere 
${\Bbb{S}}^2 = {\rm SU}(2) / {\rm U}(1)$, and each coherent
state is characterized by the unit vector
\begin{equation}
{\bf n} = (\sin\theta \cos\phi, \sin\theta \sin\phi, 
\cos\theta) .
\end{equation}
Specifically, the coherent states 
$|\Omega\rangle \equiv |j;{\bf n}\rangle$ are given by
the action of the group element 
\begin{equation}
g(\Omega) = g(\theta,\phi) = 
e^{- i \phi J_{z} } e^{- i \theta J_{y} }
\label{oSU2}
\end{equation}
on the highest-weight state $|j,j\rangle$:
\begin{eqnarray}
|j;{\bf n}\rangle & = & |j;\theta,\phi\rangle =
g(\theta,\phi) |j,j\rangle \nonumber \\
& = & \sum_{\mu=-j}^{j} {2j \choose j+\mu}^{1/2}
\cos^{j+\mu}(\theta/2) \sin^{j-\mu}(\theta/2) \nonumber \\
& & \times e^{- i \mu \phi} |j,\mu\rangle .
\label{jcohstates}
\end{eqnarray}
The invariant measure is 
\begin{equation}
 d \mu(\Omega) \equiv \frac{2j+1}{4\pi} d {\bf n} =
\frac{2j+1}{4\pi} \sin\theta\, d \theta\, d \phi , 
\end{equation}
and the corresponding delta function is 
\begin{eqnarray}
\delta(\Omega - \Omega') & \equiv & 
\frac{4\pi}{2j+1}\, \delta({\bf n} - {\bf n}') \nonumber \\
& = & \frac{4\pi}{2j+1}\, \delta(\cos\theta - \cos\theta')\,
\delta(\phi-\phi') . 
\end{eqnarray}

The harmonic functions on ${\Bbb{S}}^{2}$
are the familiar spherical harmonics: 
\begin{equation}
Y_{\nu}(\Omega) 
\equiv \sqrt{\frac{4\pi}{2j+1}}\, Y_{lm}(\theta,\phi) . 
\label{Ytp}
\end{equation}
In this context $\nu$ is the double discrete index $\{l,m\}$ 
with $l=0,1,2,\ldots$  and $m=l,l-1,\ldots,-l$. 
The transformation rule for the spherical harmonics reads
\begin{equation}
g(\alpha,\beta,\gamma) Y_{lm}(\theta,\phi) = \sum_{m'=-l}^{l}
{\cal D}_{m'm}^{(l)}(\alpha,\beta,\gamma) Y_{lm'}(\theta,\phi) ,
\label{Ylmtransform}
\end{equation}
where 
\begin{equation}
{\cal D}_{m'm}^{(l)}(\alpha,\beta,\gamma) =
\langle l,m' | g(\alpha,\beta,\gamma) | l,m \rangle
\end{equation}
is the matrix representation of SU(2) elements and 
$g(\alpha,\beta,\gamma)$ is given by Eq.\ (\ref{gEuler}).
Another property of the spherical harmonics is
\begin{equation}
Y_{lm}^{\ast}(\theta,\phi) = (-1)^{m} Y_{l,-m}(\theta,\phi)
\end{equation}

The invariant coefficients can be found using the following
expansion \cite{VaGB89}:
\begin{eqnarray}
|\langle j,{\bf n}|j,{\bf n}' \rangle|^{2} & = &
\left( \frac{1+{\bf n}\cdot{\bf n}'}{2} \right)^{2j}
\nonumber \\
& = & \sum_{l=0}^{2j} \frac{2l+1}{2j+1} 
\langle j,j;l,0|j,j \rangle^{2} 
P_{l}({\bf n}\cdot{\bf n}') ,
\label{nnform1}
\end{eqnarray}
where $P_{l}(x)$ are the Legendre polynomials and
\begin{equation}
\langle j_{1},m_{1};j_{2},m_{2}|j,m \rangle \equiv
C_{m_{1} m_{2} m}^{j_{1} j_{2} j}
\end{equation}
are the Clebsch-Gordan coefficients. Using the addition
formula for the spherical harmonics,
\begin{equation}
\frac{2l+1}{4\pi} P_{l}({\bf n}\cdot{\bf n}')
= \sum_{m=-l}^{l} Y_{lm}^{\ast}({\bf n})
Y_{lm}({\bf n}') ,
\end{equation}
Eq.\ (\ref{nnform1}) can be rewritten as
\begin{eqnarray}
|\langle j,{\bf n}|j,{\bf n}' \rangle|^{2} & = &
\frac{4\pi}{2j+1} \sum_{l=0}^{2j} \sum_{m=-l}^{l}
\langle j,j;l,0|j,j \rangle^{2} \nonumber \\
& & \times Y_{lm}^{\ast}({\bf n}) Y_{lm}({\bf n}') .
\end{eqnarray}
Comparing this result with the general formula
(\ref{ooexpansion}), we readily find that the invariant 
coefficients are given by
\begin{equation}
\tau_{\nu} \equiv \tau_{l} = \langle j,j;l,0|j,j \rangle^{2} 
= \frac{ (2j+1) [(2j)!]^{2} }{ (2j+l+1)! (2j-l)! } .
\label{taul}
\end{equation}
Note that $\tau_{l} = 0$ for $l > 2j$. The invariance
of $\tau_{l}$ is ensured by the fact that they
are independent of $m$.

The tensor operators for spin are the well-known Fano 
multipole operators \cite{Fano}, which can be written in 
the form
\begin{equation}
D_{lm} = \sqrt{\frac{2l+1}{2j+1}} \sum_{k,q=-j}^{j}
\langle j,k;l,m|j,q \rangle |j,q \rangle \langle j,k| .
\label{Dlm}
\end{equation}
Substituting expressions (\ref{Ytp}), (\ref{taul}), and 
(\ref{Dlm}) into the general formula (\ref{kerneldeff}),
we find that the SW kernel for spin is given by
\begin{eqnarray}
\Delta(\theta,\phi;s) & = & \sqrt{\frac{4\pi}{2j+1}}
\sum_{l=0}^{2j} \langle j,j;l,0|j,j \rangle^{-s} \nonumber \\
& & \times \sum_{m=-l}^{l} D_{lm} Y_{lm}^{\ast}(\theta,\phi) ,
\end{eqnarray}
which coincides for $s=0,\pm 1$ with the results by Agarwal 
\cite{Agar81} and by V\'{a}rilly and Gracia-Bond\'{\i}a 
\cite{VaGB89}.

\section{Reconstruction of quantum states}

\subsection{Basic systems and methods}

A great amount of work has been devoted in the last few years 
to the problem of determining the quantum state from information 
obtained by a set of measurements performed on an ensemble of 
identically prepared systems. The task is to reconstruct the 
density matrix $\rho$ which, according to the principles of 
quantum physics, contains all available information about the 
state of a system. Of course, the question arises which set of 
measurements provides information sufficient for the state 
reconstruction. This question first appeared in early works by 
Fano \cite{Fano57} and Pauli \cite{Pauli58} and was discussed in 
a number of papers 
\cite{Rec:early1,BaPa,Woot,Prug77,Busch91,HeSch95}. 

Recently, significant theoretical and experimental progress has 
been achieved in the reconstruction of quantum states of the
light field (see, e.g., a recent book \cite{Leon:book}).
One of the most successful reconstruction methods in this context
is the optical homodyne tomography. A tomographic approach to the
Wigner function was discussed by Bertrand and Bertrand 
\cite{BeBe87} and a quantum-optical scheme was proposed by Vogel
and Risken \cite{VoRi89}. The reconstruction of quantum states of
the light field by means of homodyne tomography was realized 
in a series of intriguing experiments \cite{Raymer,Mlynek}. 
Various methods for data analysis in optical homodyne 
tomography measurements were recently discussed 
\cite{DAMP94,KWV94,LPDA95,LMKRR96,Wun97}. The tomographic schemes 
were also generalized for the reconstruction of the joint density
matrix for two- and multi-mode optical fields 
\cite{KWV95,RMAL96,OWV97,Richter97,PTKJ97}. 
Among other approaches to the reconstruction of quantum states
of light we would like to mention the symplectic tomography
\cite{symptom} and the photon counting methods 
\cite{BaWo96,WaVo96,OpWe97}
(also known as the photon number tomography \cite{MTM97}). 
 
In the case of a single-mode microwave field inside a high-$Q$ 
cavity, a direct measurement on the system itself is impossible.
Instead, one can probe the state of the intra-cavity field via 
the detection of atoms after their interaction with the field 
mode \cite{BMS95,LuDa97,BAKW98}. Similar ideas were also 
applied to the reconstruction of the quantum motional state
of a laser-cooled ion trapped in a harmonic potential 
\cite{LuDa97,WaVo,PWCZB96,DHMi96,BLSS96,Frey97},
including a beautiful experimental realization \cite{Leibfr}.

State reconstruction procedures were proposed for various
quantum systems, for example, 
one-dimensional wave packets \cite{Royer,1dwp}, 
harmonic and anharmonic molecular vibrations 
\cite{anharm,DOZ98}, 
motional states of atom beams \cite{atbeams}, 
Bose-Einstein condensates \cite{beccond},
cyclotron states of a trapped electron \cite{MaTo97},
atomic Rydberg wave functions \cite{ChYe97}, etc.
State reconstruction methods for systems with a 
finite-dimensional state space (e.g., for spin) were 
also discussed \cite{Woot87,Leonh,BaPa,Woot,DoMa97,Agar98}.
Experimental reconstructions were also reported for 
electronic angular-momentum states of hydrogen \cite{ACBWR90} 
and for vibrational quantum states of a diatomic molecule
\cite{molvibr}.

\subsection{Displaced projectors}

It turns out that the majority of schemes used for the 
reconstruction of quantum states are related to the phase-space 
formalism. Frequently, the $Q$ function, the Wigner function, 
or other phase-space QPDs representing the density matrix $\rho$
of the system can be either measured directly or deduced in some
way from measured data. In particular, in many proposed and
realized schemes the measured quantity is the expectation
value
\begin{equation}
p_{u}(\lambda) = \langle \Gamma_{u}(\lambda) \rangle
= {\rm Tr}\, [ \rho \Gamma_{u}(\lambda) ]
\label{evalue}
\end{equation}
of a self-adjoint operator 
\begin{equation}
\Gamma_{u}(\lambda) = U(\lambda) |u \rangle\langle u|
U^{\dagger}(\lambda) ,
\end{equation}
which is a transformed projector on a quantum state 
$|u\rangle$. The unitary operator $U(\lambda)$ represents
the corresponding transformation, and the measurements are made
for a range of values of the transformation parameter $\lambda$.

We will distinguish here between two possibilities.
If $U(\lambda) = T(\Omega)$ is the phase-space 
displacement operator which represents an element of $X=G/H$, 
with $G$ being the dynamical symmetry group of given quantum 
system, we will call the observable 
$\Gamma_{u}(\lambda) = \Gamma_{u}(\Omega)$ the 
properly transformed projector (or the displaced projector). 
Otherwise $\Gamma_{u}(\lambda)$ will be called 
the improperly transformed projector.

In order to illustrate these definitions, let us consider
a quantum harmonic oscillator which is the model system for
a single mode of the quantized radiation field, 
a laser-cooled ion moving in a harmonic trap, 
or a harmonic vibrational mode of a diatomic molecule.
The corresponding symmetry group is the Heisenberg-Weyl group
H$_{3}$, and the phase space is the complex plane 
${\Bbb{C}} = {\rm H}_3 / {\rm U}(1)$ 
(see section \ref{HWsection}). In this context 
$U(\lambda) = D(\alpha)$ is the Glauber displacement operator,
and the expectation value of the displaced projector, 
\begin{equation}
p_{u}(\alpha) = {\rm Tr}\, [ \rho \Gamma_{u}(\alpha) ]
= {\rm Tr}\, [ \rho D(\alpha) \rho_{u} D^{\dagger}(\alpha) ] ,
\end{equation}
is called the operational phase-space probability distribution
\cite{Wod84,BKK95,Ban98}. Here, $\rho$ is the density matrix of 
the quantum state of the system and $\rho_{u}$ is the density 
matrix (given by the projector $|u \rangle \langle u|$ for a 
pure state) of the so-called ``quantum ruler'' state
which characterizes the measurement device. 
For example, displacing the state of the oscillator,
\begin{equation}
\rho \rightarrow \rho(\alpha) = 
D^{\dagger}(\alpha) \rho D(\alpha) ,
\hspace{0.5cm} \alpha \in {\Bbb{C}} ,
\label{rhodisplaced}
\end{equation}
and measuring the probability of finding it in the Fock
state $|n\rangle$, one obtains the operational phase-space 
probability distribution,
\begin{equation}
p_{n}(\alpha) = \langle n | \rho(\alpha) | n \rangle =
{\rm Tr}\, [ \rho \Gamma_{n}(\alpha) ] .
\end{equation}
The displaced projector
\begin{equation}
\Gamma_{n}(\alpha) = D(\alpha) |n \rangle\langle n| 
D^{\dagger}(\alpha)
\end{equation}
is obtained for $|u\rangle = |n\rangle$ being 
the Fock state. In particular, measuring the probability
to find the displaced oscillator in the ground state
$|0 \rangle$, one obtains the Husimi function 
$Q(\alpha) = \langle \alpha | \rho | \alpha \rangle$.
On the other hand, if one knows the functions $p_{n}(\alpha)$
for all values of $n$, then the Wigner function can be built
as \cite{CaGl69}
\begin{equation}
W(\alpha) = 2 \sum_{n=0}^{\infty} (-1)^{n} p_{n}(\alpha) .
\end{equation}
This formula can be generalized for QPDs with other values 
of $s$ \cite{MCKn93}:
\begin{equation}
F_{\rho}(\alpha;s) \equiv P(\alpha;s) =  \frac{2}{1-s} 
\sum_{n=0}^{\infty} \left( \frac{s+1}{s-1} \right)^{n} 
p_{n}(\alpha) .
\label{mcformula}
\end{equation}

These methods for determining the Husimi function and the Wigner 
function (and thus reconstructing the quantum state of the 
system) were first discussed by Royer \cite{Royer} in 1985. 
Recently, such a scheme for measuring the $Q$ function was 
proposed in the context of trapped ions \cite{PWCZB96}. 
Another method for the reconstruction of the motional state
of a trapped ion, proposed and experimentally realized by the
NIST group \cite{Leibfr}, employs the interaction between
the vibrational mode of the ion and its internal electronic
levels. The initial motional state is displaced in the
phase space, as in Eq.\ (\ref{rhodisplaced}), and then the 
interaction with the two-level internal subsystem is induced
for a time $t$. The population $P_{\downarrow}(t,\alpha)$ of 
the lower internal state $|\downarrow\rangle$ is measured 
for different values of displacement amplitude $\alpha$
and time $t$ (this measurement can be made with great accuracy
by monitoring the fluorescence produced in driving a resonant 
dipole transition). If $|\downarrow\rangle$ is the internal 
state at $t=0$, the signal averaged over many measurements is
\begin{equation}
P_{\downarrow}(t,\alpha) = \frac{1}{2} \left[ 1 + 
\sum_{n=0}^{\infty} p_{n}(\alpha) \cos(2 \Omega_{n,n+1} t)
e^{-\gamma_{n} t} \right] ,
\end{equation}
where $\Omega_{n,n+1}$ are the Rabi frequencies and $\gamma_{n}$ 
are the experimentally determined decay constants. This
relation allows to determine the populations $p_{n}(\alpha)$ of
the displaced motional eigenstates. As one can see from
Eq.~(\ref{mcformula}), the functions $p_{n}(\alpha)$ in their 
turn can be used to calculate the QPDs $P(\alpha;s)$ (e.g., the
Wigner function). Alternatively, the density matrix in the
Fock representation can be deduced directly from $p_{n}(\alpha)$.

In the optical domain, the function $p_{n}(\alpha)$ can be 
determined in principle as the probability of recording $n$ counts 
with an ideal photodetector exposed to the displaced light field.
In practice, one could use the unbalanced homodyning detection
\cite{BaWo96,WaVo96,OpWe97,MTM97}, in which the signal field
is mixed in a beam splitter with the local oscillator of
coherent amplitude $\beta$ and the photon statistics of the 
superimposed field is then counted by a photodetector of 
quantum efficiency $\zeta$. The resulting counting statistics 
is denoted by $p_{n}(\alpha,\eta)$, where $\alpha = -R\beta/T$
is the effective displacement amplitude ($T$ and $R$ are the 
transmission and reflection coefficients of the beam splitter)
and $\eta = \zeta |T|^{2}$ is the overall quantum efficiency. 
In this realistic situation formula (\ref{mcformula}) should
be replaced by the following result \cite{WaVo96}:
\begin{equation}
P(\alpha;s) =  \frac{2}{1-s} \sum_{n=0}^{\infty} 
\left[ \frac{2+\eta(s-1)}{\eta(s-1)} \right]^{n} 
p_{n}(\alpha,\eta) .
\end{equation}
This method of state reconstruction is sometimes called the
photon number tomography.

As an example of measurements with improperly transformed
projectors, we mention the optical homodyne tomography
\cite{VoRi89,Raymer} in which one measures the probability 
distribution $P(x,\theta)$ for the rotated field quadrature
\begin{equation}
x_{\theta} = x \cos\theta + p \sin\theta =
U(\theta) x U^{\dagger}(\theta) .
\end{equation}
The field quadratures $x$ and $p$ can be viewed as the scaled 
position and momentum operators of the harmonic oscillator, 
with $a = 2^{-1/2}(x + i p)$, and
$U(\theta) = \exp(i \theta a^{\dagger} a)$ is the 
rotation operator (known in optics as the phase shifter)
on the phase plane. $U(\theta)$ represents an element
of the SO(2)$\sim$U(1) subgroup of the oscillator group 
H$_{4}$ whose algebra is spanned by 
$\{ I,a,a^{\dagger}, a^{\dagger} a \}$.
The improperly transformed projector is given by
\begin{equation}
\Gamma_{x}(\theta) = U(\theta) |x \rangle\langle x|
U^{\dagger}(\theta) ,
\end{equation}
where $|x\rangle$ are the position eigenstates.
The measured distribution $P(x,\theta)$ can be used for
determining the Wigner function via the inverse Radon 
transform \cite{BeBe87,VoRi89,Raymer}. 
Alternatively, the density matrix in some basis (e.g., in the 
Fock basis) can be deduced directly from $P(x,\theta)$ by
averaging a set of pattern functions 
\cite{DAMP94,KWV94,LPDA95,LMKRR96}. Another example of
measurements with improperly transformed projectors is the
symplectic tomography \cite{symptom}, in which the phase-space
rotation is accompanied by the squeezing transformation.

In the case of measurements with improperly transformed 
projectors, a general group-theoretic approach is problematic,
because the number of possible transformations is very large
and one should consider each situation separately.
On the other hand, the method of properly transformed 
projectors works uniformly for physical systems with 
different symmetry groups. For example, in the case of the 
SU(2) symmetry (e.g., spin, two-level atoms, etc.), proposals 
appeared \cite{DoMa97,Agar98} for measuring the $Q$ 
function, 
\begin{equation}
Q({\bf n}) = \langle j,{\bf n}| \rho |j,{\bf n} \rangle
= {\rm Tr}\, (\rho |j,{\bf n} \rangle\langle j,{\bf n}| ) ,
\end{equation}
or, more generally, for measuring the probability 
\begin{eqnarray}
& p_{\mu}({\bf n}) = {\rm Tr}\, 
[\rho \Gamma_{\mu}({\bf n}) ] , & \\
& \Gamma_{\mu}({\bf n}) = g({\bf n}) 
|j,\mu \rangle\langle j,\mu| g^{-1}({\bf n}) 
\end{eqnarray}
of finding the displaced system 
\begin{equation}
\rho({\bf n}) = g^{-1}({\bf n}) \rho g({\bf n}) ,
\hspace{0.5cm} {\bf n} \in {\Bbb{S}}^{2} 
\end{equation}
in the state $|j,\mu\rangle$. These 
ideas for spin are conceptually very similar to the proposals 
in the context of optical fields or trapped ions. Therefore, 
it seems natural to apply the phase-space formalism developed 
above to the general group-theoretic description of the
state reconstruction method based on the measurement of 
displaced projectors.

\subsection{General reconstruction formalism}

From the practical point of view, the reconstruction procedure 
consists of two steps. First, the system described by the density 
matrix $\rho$ is displaced in the phase space:
\begin{equation}
\rho \rightarrow \rho(\Omega) = T^{-1}(\Omega) \rho T(\Omega) ,
\hspace{0.5cm} \Omega \in X .
\end{equation}
The second step is the measurement of the probability to 
find the (displaced) system in a quantum state $|u\rangle$,
\begin{equation}
p_{u}(\Omega) = \langle u|\rho(\Omega)|u \rangle .
\end{equation}
Repeating this procedure for a large number of phase-space
points $\Omega$, one can in principle determine the function
$p_{u}(\Omega)$. 

\subsubsection{More about displaced projectors}

The information contained in the function $p_{u}(\Omega)$ is
enough for the reconstruction of the density matrix. It is
convenient to analyze this problem with the help of the
displaced projector,
\begin{equation}
\Gamma_{u}(\Omega) = T(\Omega) |u \rangle\langle u| 
T^{-1}(\Omega) ,
\end{equation}
whose expectation value gives the measured function 
$p_{u}(\Omega)$, as in Eq.\ (\ref{evalue}). 
The displaced projector satisfies a number of useful 
properties:
\begin{enumerate}
\item[(i)] It is a self-adjoint operator,
\begin{equation}
\Gamma_{u}^{\dagger}(\Omega) = \Gamma_{u}(\Omega)
\hspace{0.5cm} \forall \Omega \in X .
\end{equation}
Since $p_{u}(\Omega)$ is not only real but also non-negative
(this is a probability), $\Gamma_{u}(\Omega)$ is also a
non-negatively defined operator.
\item[(ii)] 
Provided that the state $|u \rangle$ is normalized, 
$\Gamma_{u}(\Omega)$ is a trace-class operator of trace one, 
and the following standardization condition holds,
\begin{equation}
\int_{X} d \mu(\Omega) \Gamma_{u}(\Omega) = I .
\end{equation}
This implies the normalization of $p_{u}(\Omega)$,
\begin{equation}
\int_{X} d \mu(\Omega) p_{u}(\Omega) = 1 .
\end{equation}
\item[(iii)]
The displaced projector is manifestly covariant,
\begin{equation}
T(g) \Gamma_{u}(\Omega) T(g^{-1}) = \Gamma_{u}(g\cdot\Omega) .
\end{equation}
Consequently, if $p_{u}(\Omega)$ corresponds to the initial
density matrix $\rho$, the function $p_{u}(g\cdot\Omega)$
will correspond to the transformed density matrix 
$\rho(g) = T(g^{-1}) \rho T(g)$.
\end{enumerate}

Denoting the density matrix of the quantum ruler state by 
$\rho_{u}$ (which is $|u \rangle \langle u|$ for a pure
state), the operational phase-space probability distribution
reads
\begin{equation}
p_{u}(\Omega) = {\rm Tr}\, [ \rho T(\Omega) \rho_{u} 
T^{-1}(\Omega) ] .
\end{equation}
Using the inverse Weyl rule (\ref{invmap}) for the density
matrix $\rho$, we obtain
\begin{displaymath}
p_{u}(\Omega) = \int_{X} d \mu(\Omega') P(\Omega';s)
{\rm Tr}\, [ T^{-1}(\Omega) \Delta(\Omega';-s) T(\Omega) 
\rho_{u} ] ,
\end{displaymath}
where $P(\Omega;s) \equiv F_{\rho}(\Omega;s)$ is the SW symbol 
of $\rho$. Now, the covariance property (\ref{covar1}) can be 
used to obtain the following expression:
\begin{equation}
p_{u}(\Omega) = \int_{X} d \mu(\Omega') 
P(\Omega \cdot \Omega';s) P_{u} (\Omega';-s) ,
\label{puconvol}
\end{equation}
where $P_{u} (\Omega;s)$ is the SW symbol of $\rho_{u}$.
Therefore, the operational phase-space probability 
distribution $p_{u}(\Omega)$ is given by a convolution of the 
two QPDs representing the quantum state of the system and the 
quantum ruler state of the measurement apparatus.
In the particular case of the Heisenberg-Weyl group and for
$s=0$, the general expression (\ref{puconvol}) reduces to
the known result \cite{BKK95}
\begin{equation}
p_{u}(\alpha) = \int_{ {\Bbb{C}} } \frac{ d^{2} \alpha' }{\pi}
W(\alpha+\alpha') W_{u}(\alpha') .
\end{equation}

If the quantum ruler state $|u \rangle = |\psi_{0} \rangle$ 
is the reference state of the coherent-state basis, then 
\begin{equation}
\Gamma_{\psi_{0}}(\Omega) = |\Omega\rangle\langle\Omega|
= \Delta(\Omega;-1) 
\end{equation}
is the SW kernel with $s=-1$, and 
\begin{equation}
p_{\psi_{0}}(\Omega) = \langle\Omega| \rho |\Omega\rangle
= Q_{\rho}(\Omega)
\end{equation}
is the $Q$ function. However, except for this coincidence,
the displaced projectors are not the SW kernels, as they
do not satisfy the traciality condition. On the other hand,
the functions $p_{u}(\Omega)$ differ from the majority of
QPDs, as they are positive on the whole phase space (which
reflects the fact that they are measurable probabilities).
Usually the state $|u \rangle$ is chosen to belong to some
complete orthonormal basis $\{ |\phi_{n}\rangle \}$ which 
consists of energy eigenstates of a natural Hamiltonian
of the physical system (e.g., the Fock basis for a harmonic
oscillator or $J_{z}$ eigenstates for spin).
Then there exists the relation
\begin{equation}
\sum_{n} p_{\phi_{n}}(\Omega) = 1 , 
\end{equation}
which follows from the completeness of the basis.

\subsubsection{Entropy}

A useful quantity for analyzing statistical properties of
the system (in particular, the amount of noise) is the 
entropy. A phase-space version of the entropy can be
introduced in the following way,
\begin{equation}
S_{u} = - \int_{X} d \mu(\Omega) p_{u}(\Omega) 
\ln p_{u}(\Omega) .
\label{entropy}
\end{equation}
For $|u \rangle = |\psi_{0} \rangle$, Eq.\ (\ref{entropy})
gives
\begin{equation}
S = - \int_{X} d \mu(\Omega)\, Q_{\rho}(\Omega) 
\ln Q_{\rho}(\Omega) ,
\end{equation}
which is a generalization of the Wehrl entropy \cite{Wehrl} 
that was defined originally on the flat phase space of the 
Weyl-Heisenberg group. The entropy (\ref{entropy}) can be
useful in the reconstruction procedure, as it is a sensitive
measure of the noise added to the system during the 
displacement and detection processes. A similar situation
exists also in the field of signal processing \cite{Trees68}. 
There $|u \rangle$ represents the test signal and 
$p_{u}(\Omega)$ is a distribution on the time-frequency space. 
One can produce various test signals $|u \rangle$ and compute 
the corresponding entropies $S_{u}$. Choosing $|u \rangle$
that minimizes the entropy, one obtains the optimal form
of pattern analysis (in particular, this method allows
to achieve data compression).

\subsubsection{Harmonic expansions}

A useful expression for $p_{u}(\Omega)$ can be derived in the
following way. Using the expansion
\begin{equation}
\rho = \sum_{\nu} {\cal R}_{\nu} D_{\nu} ,
\hspace{0.8cm} {\cal R}_{\nu} \equiv {\rm Tr}\, 
(\rho D_{\nu}^{\dagger}) ,
\end{equation}
we obtain
\begin{eqnarray}
p_{u}(\Omega) & = & \sum_{\nu} {\cal R}_{\nu} \langle u| 
T^{-1}(\Omega) D_{\nu} T(\Omega) |u \rangle \nonumber \\
& = & \sum_{\nu} {\cal R}_{\nu} 
\sum_{\nu'} T_{\nu'\nu}^{-1}(\Omega) 
\langle u| D_{\nu'} |u \rangle .
\end{eqnarray}
Expanding $p_{u}(\Omega)$ in the basis of harmonic functions,
\begin{equation}
p_{u}(\Omega) = \sum_{\nu} \kappa_{\nu}^{(u)} {\cal R}_{\nu} 
Y_{\nu}(\Omega) ,
\label{pnexpans}
\end{equation}
we identify the coefficients $\kappa_{\nu}^{(u)}$ by means 
of the following relation
\begin{equation}
\kappa_{\nu}^{(u)} Y_{\nu}(\Omega) = \langle u| 
T(\Omega^{-1}) D_{\nu} T(\Omega) |u \rangle .
\label{kY} 
\end{equation}
Formally, we can write
\begin{eqnarray}
\kappa_{\nu}^{(u)} & = & \tau_{\nu}^{-1/2} 
\int_{X} \int_{X} d \mu(\Omega) d \mu(\Omega') 
Y_{\nu}^{\ast}(\Omega) \nonumber \\
& & \times Y_{\nu}(\Omega\cdot\Omega') 
|\langle u|\Omega'\rangle|^{2} ,
\end{eqnarray}
but actually Eq.\ (\ref{kY}) is more convenient for
calculating the coefficients $\kappa_{\nu}^{(u)}$.

Equation (\ref{pnexpans}) for the functions $p_{u}(\Omega)$ 
corresponds to the expansion
\begin{equation}
\Gamma_{u}(\Omega) = \sum_{\nu} \kappa_{\nu}^{(u)} 
Y_{\nu}^{\ast}(\Omega) D_{\nu} = 
\sum_{\nu} \kappa_{\nu}^{(u)} 
Y_{\nu}(\Omega) D_{\nu}^{\dagger} 
\label{gammaexpans}
\end{equation}
for the displaced projectors. It follows from the properties
of $\Gamma_{u}(\Omega)$ that the coefficients 
$\kappa_{\nu}^{(u)}$ are positive and possess the same
invariance properties as $\tau_{\nu}$.
Using the general result (\ref{FHtrans}), we obtain the
relation between the functions $p_{u}(\Omega)$ and
$p_{v}(\Omega)$, corresponding to different quantum ruler
states $|u\rangle$ and $|v\rangle$,
\begin{eqnarray}
& & p_{u}(\Omega) = \int_{X} d \mu(\Omega') 
K_{uv}(\Omega,\Omega') p_{v}(\Omega') ,  \\
& & K_{uv}(\Omega,\Omega') = \sum_{\nu} 
\frac{\kappa_{\nu}^{(u)}}{\kappa_{\nu}^{(v)}}
Y_{\nu}(\Omega) Y_{\nu}^{\ast}(\Omega') .
\end{eqnarray}

\subsubsection{Deducing the density matrix and 
quasiprobabilities} 

Knowledge of the phase-space function $p_{u}(\Omega)$ 
allows to reconstruct the density matrix in a simple way:
\begin{equation}
{\cal R}_{\nu} = \left[ \kappa_{\nu}^{(u)} \right]^{-1}
\int_{X} d \mu(\Omega) p_{u}(\Omega) Y_{\nu}^{\ast}(\Omega) .
\label{Rreconstr}
\end{equation}
Formally, we can also represent the density matrix by means
of an integral transform of the displaced projector:
\begin{equation}
\rho = \int_{X} d \mu(\Omega) r_{u}(\Omega) \Gamma_{u}(\Omega) .
\end{equation}
This relation gives the density matrix in terms of a
phase-space function $r_{u}(\Omega)$, and in this sense it is 
the inverse of Eq.\ (\ref{evalue}).
The function $r_{u}(\Omega)$ is defined by its harmonic 
expansion,
\begin{equation}
r_{u}(\Omega) = \sum_{\nu} [\kappa_{\nu}^{(u)}]^{-1} 
{\cal R}_{\nu} Y_{\nu}(\Omega) .
\label{rfunction}
\end{equation}
We also obtain the following relation between the functions
$r_{u}(\Omega)$ and $p_{u}(\Omega)$,
\begin{equation}
p_{u}(\Omega) = \int_{X} d \mu(\Omega') r_{u}(\Omega')
{\rm Tr}\,[ \Gamma_{u}(\Omega) \Gamma_{u}(\Omega') ] ,
\end{equation}
where
\begin{eqnarray}
{\rm Tr}\,[ \Gamma_{u}(\Omega) \Gamma_{v}(\Omega') ] & = &
| \langle u| T^{-1}(\Omega) T(\Omega') |v \rangle |^{2} 
\nonumber \\
& = & \sum_{\nu} \kappa_{\nu}^{(u)} \kappa_{\nu}^{(v)} 
Y_{\nu}(\Omega) Y_{\nu}^{\ast}(\Omega') .
\end{eqnarray}
Certainly, the most convenient way for deducing the density
matrix from the measured functions $p_{u}(\Omega)$ is by
calculating the coefficients ${\cal R}_{\nu}$ via Eq.\
(\ref{Rreconstr}).

The measured functions $p_{u}(\Omega)$ can be used also
for the reconstruction of various QPDs which represent the
density matrix in the phase-space formulation.
According to the general expression (\ref{ffform}), the 
QPDs for the density matrix $\rho$ are given by the harmonic
expansion
\begin{equation}
F_{\rho}(\Omega;s) \equiv P(\Omega;s) = \sum_{\nu} 
\tau_{\nu}^{-s/2} {\cal R}_{\nu} Y_{\nu}(\Omega) .
\label{Poexpans}
\end{equation}
Therefore, one can just use the coefficients 
${\cal R}_{\nu}$ calculated via Eq.\ (\ref{Rreconstr}).
On the other hand, Eq.\ (\ref{FHtrans}) can be used to obtain 
the relation between the QPDs $P(\Omega;s)$ and the measured
functions $p_{u}(\Omega)$:
\begin{eqnarray}
& & P(\Omega;s) = \int_{X} d \mu(\Omega') 
K_{us}^{-}(\Omega,\Omega') p_{u}(\Omega') ,  
\label{Kminusrel} \\
& & p_{u}(\Omega) = \int_{X} d \mu(\Omega') 
K_{us}^{+}(\Omega,\Omega') P(\Omega';s) ,
\label{Kplusrel}
\end{eqnarray}
where the transformation kernels are
\begin{equation}
K_{us}^{\pm}(\Omega,\Omega') = \sum_{\nu} 
\left[ \kappa_{\nu}^{(u)} \tau_{\nu}^{s/2} \right]^{\pm 1}
Y_{\nu}(\Omega) Y_{\nu}^{\ast}(\Omega') .
\end{equation}
It can be easily shown that $K_{us}^{+}(\Omega,\Omega') = 
P_{u}(\Omega^{-1} \cdot \Omega';-s)$ where $P_{u}(\Omega;s)$ 
is the SW symbol of $|u\rangle \langle u|$, so 
Eq.~(\ref{Kplusrel}) is consistent with the relation
(\ref{puconvol}).

As was already mentioned, if the state $|u\rangle$ is the
reference state $|\psi_{0}\rangle$ of the coherent-state 
basis, then 
$\Gamma_{\psi_{0}}(\Omega) = |\Omega\rangle\langle\Omega|$
and the measured function $p_{\psi_{0}}(\Omega)$ coincides 
with the function $Q_{\rho}(\Omega) = P(\Omega;-1)$.
Comparing the harmonic expansions (\ref{pnexpans}) and 
(\ref{Poexpans}) for the case $|u\rangle = |\psi_{0}\rangle$, 
we find the following relation:
\begin{equation}
\kappa_{\nu}^{(\psi_{0})} = \tau_{\nu}^{1/2} .
\end{equation}
In this case we also obtain that the function 
$r_{\psi_{0}}(\Omega)$ of Eq.~(\ref{rfunction}) is just the 
$P$ function,
\begin{equation}
r_{\psi_{0}}(\Omega) = P_{\rho}(\Omega) = P(\Omega;1) .
\end{equation}
Note that in the case of the Heisenberg-Weyl group one can 
also calculate the QPDs using the formula (\ref{mcformula}). 

\subsubsection{Examples}

We see that the mathematical procedure of the reconstruction 
of the density matrix $\rho$ and its QPDs $P(\Omega;s)$ from 
the measured probability $p_{u}(\Omega)$ actually consists of 
the simple transformation (\ref{Rreconstr}). The mathematical 
tools one needs for this procedure are the harmonic functions 
$Y_{\nu}(\Omega)$ and the invariant coefficients $\tau_{\nu}$ 
and $\kappa_{\nu}^{(u)}$. In what follows we compute the 
explicit form of $\kappa_{\nu}^{(u)}$ for simple but 
instructive examples of the Heisenberg-Weyl group (with
$|u\rangle$ being a Fock state) and the SU(2) group
(with $|u\rangle$ being a $J_{z}$ eigenstate).

In the case of the Heisenberg-Weyl group, we consider the
probability $p_{n}(\alpha)$ to find the displaced initial
state in the Fock state $|n\rangle$ ($n=0,1,2,\ldots$).
Then Eq.\ (\ref{kY}) can be rewritten in the form
\begin{equation}
\kappa^{(n)}(\xi) Y(\xi,\alpha) = \langle n| D(-\alpha)
D(\xi) D(\alpha) |n \rangle .
\end{equation}
Using Eq.\ (\ref{DDDidentity}), we obtain
\begin{equation}
D(-\alpha) D(\xi) D(\alpha) = Y(\xi,\alpha) D(\xi) .
\end{equation}
Therefore, the $\kappa$ coefficients are given by
\begin{equation}
\kappa^{(n)}(\xi) = \langle n| D(\xi) |n \rangle =
\exp(-\mbox{$\frac{1}{2}$} |\xi|^{2}) L_{n}(|\xi|^{2}) .
\end{equation}
Of course, for $n=0$ one gets 
$\kappa^{(0)}(\xi) = [\tau(\xi)]^{1/2}$.

In the case of the SU(2) group, we consider the
probability $p_{\mu}(\theta,\phi)$ to find the displaced initial
state in the $J_{z}$ eigenstate $|j,\mu\rangle$ 
($\mu=j,j-1,\ldots,-j$). Then Eq.\ (\ref{kY}) takes the
following form
\begin{equation}
\kappa^{(\mu)}_{l m} Y_{l m}(\theta,\phi) = 
\sqrt{ \frac{2j+1}{4\pi} } \langle j,\mu| g^{-1}(\theta,\phi) 
D_{l m} g(\theta,\phi) |j,\mu \rangle .
\label{s21eq}
\end{equation}
Using the parameterization (\ref{oSU2}) for $g(\theta,\phi)$ and 
the transformation rule (\ref{Ylmtransform}), we can write
\begin{equation}
g^{-1}(\theta,\phi) D_{l m} g(\theta,\phi) = \sum_{m'=-l}^{l}
{\cal D}^{(l)}_{m' m}(0,\theta,\phi) D_{l m'} .
\end{equation}
Since the matrix element of the tensor operator,
\begin{equation}
\langle j,\mu| D_{l m'} |j,\mu \rangle = 
\sqrt{ \frac{2l+1}{2j+1} }
\langle j,\mu;l,m'|j,\mu \rangle ,
\end{equation}
vanishes unless $m' = 0$, Eq.\ (\ref{s21eq}) reads
\begin{displaymath}
\kappa^{(\mu)}_{l m} Y_{l m}(\theta,\phi) = 
\sqrt{ \frac{2l+1}{4\pi} } {\cal D}^{(l)}_{0 m}(0,\theta,\phi)
\langle j,\mu;l,0|j,\mu \rangle .
\end{displaymath}
Taking into account the fact that
\begin{equation}
{\cal D}^{(l)}_{0 m}(\alpha,\beta,\gamma) = 
\sqrt{ \frac{4\pi}{2l+1} } Y_{l m}(\beta,\gamma) ,
\end{equation}
we finally obtain that the $\kappa$ coefficients are 
independent of the index $m$:
\begin{equation}
\kappa^{(\mu)}_{l} = \langle j,\mu;l,0|j,\mu \rangle .
\end{equation}

For $\mu=j$, one finds $\kappa^{(j)}_{l} = \tau_{l}^{1/2}$. 
Indeed, according to the definition (\ref{jcohstates}) of 
the SU(2) coherent states, the function 
$Q(\theta,\phi) = \langle j,{\bf n}| \rho |j,{\bf n} \rangle$
coincides with the probability $p_{j}(\theta,\phi)$ to find
the displaced system in the highest spin state
$|j,j\rangle$. It is not difficult to see that the
probability $p_{-j}(\theta,\phi)$ to find the displaced 
system in the lowest spin state $|j,-j\rangle$ is equal
to $Q(\theta+\pi,\phi)$.
As an application of the relation (\ref{Kminusrel}), we
also obtain the following expression for the SU(2) Wigner 
function in terms of the measured probability 
$p_{\mu}({\bf n})$, 
\begin{equation}
W({\bf n}) = \sum_{l=0}^{2 j} \frac{(4\pi)^{-1} (2l+1)}{
\langle j,\mu;l,0|j,\mu \rangle } \int_{X} d {\bf n}'
P_{l}({\bf n} \cdot {\bf n}') p_{\mu}({\bf n}') ,
\end{equation}
where $P_{l}(x)$ are the Legendre polynomials.

\subsection{Informational completeness and unsharp 
measurements} 

When the question of the state reconstruction arises,
it is understood that the set of measurements one makes
on an ensemble of identically prepared systems should
give complete information about the quantum state.
In particular, if one measures expectation values of
some observables, it is natural to ask how many such
observables are needed to characterize completely
the state of the system. In this sense a set of 
observables, whose expectation values are sufficient to 
reconstruct the quantum state (or, equivalently, to 
distinguish between different states), can be considered
as informationally complete. A formal definition is as
follows \cite{Prug77}: A set of bounded operators
${\frak{A}} = \{ A \}$ on ${\cal H}$ is said to be 
informationally complete if for density matrices 
$\rho,\rho'$ the equality of expectation values,
\begin{equation}
{\rm Tr}\, (\rho A) = {\rm Tr}\, (\rho' A)
\hspace{0.5cm} \forall A \in {\frak{A}} ,
\end{equation}
implies $\rho = \rho'$. 

The informational completeness of positive operator-valued 
measures covariant with respect to Heisenberg-Weyl, affine, 
and Galilei groups was recently discussed in Ref.\ 
\cite{HeSch95}. This subject was shown \cite{HeSch95} to 
be of importance not only in quantum mechanics but also 
in signal processing where a problem exists of extracting
information from nonstationary signals and images. Another
interesting feature is that both in quantum mechanics and 
in signal processing the phase-space formulation is of
great importance for approaching this kind of problems.

It would be interesting to analyze the results of the
present paper from the point of view of informational 
completeness. First, it is evident from the expansion
\begin{equation}
\rho = \sum_{\nu} {\rm Tr}\, 
(\rho D_{\nu}^{\dagger}) D_{\nu} 
\end{equation}
that the orthonormal set $\{ D_{\nu} \}$ of the tensor 
operators is informationally complete. Correspondingly, 
the set $\{ \Delta(\Omega;s) \ | \ \Omega \in X \}$ of the 
SW kernels for each $s$ is also informationally complete. 
This fact is reflected by the inverse Weyl rule written as
\begin{equation}
\rho = \int_{X} d \mu(\Omega) {\rm Tr}\, 
[\rho \Delta(\Omega;s)] \Delta(\Omega;-s) . 
\end{equation}
In other words, the density matrix can be uniquely
reconstructed from its $s$-parameterized QPD
$P(\Omega;s) = {\rm Tr}\, [\rho \Delta(\Omega;s)]$.
From the practical point of view, the important thing
is the informational completeness of the set
$\{ \Gamma_{u}(\Omega) \ | \ \Omega \in X \}$ of the 
displaced projectors for any $|u\rangle \in {\cal H}$.
This fact was formally proved in Ref.\ \cite{HeSch95}.
Here, we presented a simple algorithm (based on
the method of harmonic expansion) for the reconstruction
of the density matrix from the measurable 
probabilities 
$p_{u}(\Omega) = {\rm Tr}\, [\rho \Gamma_{u}(\Omega)]$.
This reconstruction procedure clearly implies the
informational completeness of the set
$\{ \Gamma_{u}(\Omega) \ | \ \Omega \in X \}$.

One of the useful features of the method of displaced 
projectors is the ability to take into account in a simple 
way the unsharpness of a realistic measurement. Of course, it 
is impossible in practice to make a completely accurate
displacement to a specified point $\Omega$ on the phase space. 
For example, one should take into account the phase and 
intensity fluctuations of a classical microwave source that 
displaces the quantum state of the radiation field in a cavity, 
or instabilities of a classical driving field that displaces 
the motional state of a trapped ion. Similarly, the so-called 
coarse graining problem arises in radar analysis due to 
frequency instabilities of the test signal or uncertainties in 
timing of signal initiation.
As a result, the probabilities $p_{u}(\Omega)$ should be
integrated over the variation range. This yields the 
expectation value of the localization operator defined by 
\cite{HeSch95}
\begin{equation}
Z_{u}(f) = \int_{X} d \mu(\Omega)\, f(\Omega) 
\Gamma_{u}(\Omega) ,
\end{equation}
where $f(\Omega)$ is a localization function. $Z_{u}(f)$ has
a purely discrete spectrum, is bounded when 
$f \in L^{k}(X,\mu)$, $k \geq 1$, and is self-adjoint when $f$ 
is real. In particular, let $B$ be a region (more specifically,
a Borel set) in $X$, with the characteristic function
$\chi_{B}(\Omega)$ that equals $1$ for $\Omega \in B$ and $0$
otherwise. Taking $f(\Omega) = \chi_{B}(\Omega)$, one obtains
\begin{equation}
Z_{u}(B) = \int_{B} d \mu(\Omega)\, \Gamma_{u}(\Omega) .
\end{equation}
It is not difficult to see that the informational completeness 
of the set $\{ \Gamma_{u}(\Omega) \ | \ \Omega \in X \}$ of 
displaced projectors implies the informational completeness of 
the set $\{ Z_{u}(B) \ | \ B \in \mbox{Borel sets of } X  \}$
of localization operators. Therefore, in the case of realistic
unsharp measurements, the localization operators may be
conveniently used for analysis and reconstruction of quantum 
states or electronic signals and images.

\section{Conclusions}

In the present paper we propose a simple algorithm for
constructing the SW kernels which implement the linear 
bijective mapping between Hilbert-space operators and phase-space 
functions for physical systems possessing Lie-group symmetries. 
The constructed kernels are manifestly covariant under the 
action of the corresponding dynamical symmetry group and satisfy 
the traciality condition which ensures that quantum expectations 
can be represented by statistical-like averages over the phase 
space. Adding the noncommutative twisted product that equips 
phase-space functions with the algebraic structure of quantum 
operators, an autonomous phase-space formulation of quantum 
mechanics is developed.

It turns out that the concept of phase space naturally emerges 
in the majority of schemes proposed for the reconstruction of
quantum states as well as in the standard methods of signal 
analysis. In particular, we focus on the method based on
measurements of displaced projectors and develop its general 
group-theoretic description. We do so by applying the same 
technique of harmonic expansions on the phase space that was
used for the construction of the SW kernels. The problem of
the state reconstruction is also approached using the concept
of informational completeness, and the role of localization
operators in describing realistic measurements is discussed.

\acknowledgements

This work was supported by the Fund for Promotion of Research 
at the Technion, by the Technion VPR Fund---Promotion of 
Sponsored Research, and by the Technion VPR Fund---The Harry 
Werksman Fund.

\end{multicols}


\vspace*{-6mm}
\begin{references}
%
\bibitem[\ast]{email1} costya@physics.technion.ac.il
\bibitem[\dag]{email2} ady@physics.technion.ac.il
%
%
\bibitem{Wig32} E. Wigner, Phys. Rev. {\bf 40}, 749 (1932).
%
\bibitem{Moy49} J. E. Moyal, Proc. Cambridge Philos. Soc. {\bf 45}, 
        99 (1949).
%
\bibitem{Weyl} H. Weyl, Z. Phys. {\bf 46}, 1 (1927);
        {\em The Theory of Groups and Quantum Mechanics\/} 
        (Dover, New York, 1950).  
%
\bibitem{CaZa83} P. Carruthers and F. Zachariasen,
        Rev. Mod. Phys. {\bf 55}, 245 (1983).
\bibitem{Hill84} M. Hillery, R. F. O'Connell, M. O. Scully, and 
        E. P. Wigner, Phys. Rep. {\bf 106}, 121 (1984).
\bibitem{Kim91} Y. S. Kim and M. E. Noz, {\em Phase Space Picture 
        of Quantum Mechanics} (World Scientific, Singapore, 1991).
\bibitem{Lee95} H.-W. Lee, Phys. Rep. {\bf 259}, 147 (1995).
\bibitem{Gade95} M. Gadella, Fortschr. Phys. {\bf 43}, 229 (1995).
\bibitem{Schroek96} F. E. Schroek, Jr., {\em Quantum Mechanics
        on Phase Space\/} (Kluwer, Dordrecht, 1996).
%
\bibitem{CaGl69} K. E. Cahill and R. J. Glauber, Phys. Rev. 
        {\bf 177}, 1857, (1969); {\bf 177}, 1882 (1969).
\bibitem{AgWo70} G. S. Agarwal and E. Wolf, Phys. Rev. D {\bf 2}, 
        2161 (1970); {\bf 2}, 2187 (1970); {\bf 2}, 2206 (1970).
\bibitem{Gla63} R. J. Glauber, Phys. Rev. {\bf 131}, 2766 (1963).
\bibitem{Sudar63} E. C. G. Sudarshan, Phys. Rev. Lett. 
        {\bf 10}, 277 (1963).
\bibitem{Husimi} K. Husimi, Proc. Phys. Math. Soc. Japan
        {\bf 22}, 264 (1940).
        See also Y. J. Kano, J. Math. Phys. {\bf 6}, 1913 (1965);
        C. L. Mehta and E. C. G. Sudarshan, Phys. Rev. 
        {\bf 138}, B274 (1965).
%
\bibitem{Gardiner} C. W. Gardiner, {\em Handbook of Stochastic 
        Methods}, 2nd ed. (Springer, Berlin, 1985);
        {\em Quantum Noise} (Springer, Berlin, 1991).
\bibitem{Perina} J. Perina, {\em Quantum Statistics of Linear and 
        Non-linear Optical Phenomena}, 2nd ed. 
        (Kluwer, Dordrecht, 1991).
%
\bibitem{Leon:book} U. Leonhardt, {\em Measuring the Quantum State 
        of Light} (Cambridge Univ. Press, Cambridge, 1997).
%
\bibitem{Bay78} F. Bayen, M. Flato, C. Fronsdal, A. Lichnerowicz,
        and D. Sternheimer, Ann. Phys. (N.Y.) 
        {\bf 111}, 61 (1978); {\bf 111}, 111 (1978).
\bibitem{Fron79a} C. Fronsdal, J. Math. Phys. {\bf 20}, 2226 (1979).
\bibitem{Fron79b} C. Fronsdal, Rep. Math. Phys. {\bf 15}, 111 (1979).
\bibitem{Huynh} T. V. Huynh, Lett. Math. Phys. {\bf 4}, 201 (1980);
        J. Math. Phys. {\bf 23}, 1082 (1982).
\bibitem{BFLS84} H. Basart, M. Flato, A. Lichnerowicz, and 
        D. Sternheimer, Lett. Math. Phys. {\bf 8}, 483 (1984).
\bibitem{MoON} C. Moreno and P. Ortega-Navarro, 
        Ann. Inst. Henri Poinc. A {\bf 38}, 215 (1983);
        Lett. Math. Phys. {\bf 7}, 181 (1983); 
        C. Moreno, {\em ibid.} {\bf 12}, 217 (1986);
        {\bf 13}, 245 (1987).
%
\bibitem{Anton98} For a recent group-theoretic approach to the
        Moyal quantization, see F. Antonsen,
        Int. J. Theor. Phys. {\bf 37}, 697 (1998).
%
\bibitem{Stra56} R. L. Stratonovich, Zh. Eksp. Teor. Fiz. {\bf 31},
        1012 (1956) [Sov. Phys. JETP {\bf 4}, 891 (1957)].
%
\bibitem{GBVa88} J. M. Gracia-Bond\'{\i}a and J. C. V\'{a}rilly,
        J. Phys. A {\bf 18}, L879 (1988).
\bibitem{VaGB89} J. C. V\'{a}rilly and J. M. Gracia-Bond\'{\i}a,
        Ann. Phys. (N.Y.) {\bf 190}, 107 (1989).
\bibitem{CaGBVa90} J. F. Cari\~{n}ena, J. M. Gracia-Bond\'{\i}a, 
        and J. C. V\'{a}rilly, J. Phys. A {\bf 23}, 901 (1990).
\bibitem{FiGBVa90} H. Figueroa, J. M. Gracia-Bond\'{\i}a, and
        J. C. V\'{a}rilly, J. Math. Phys. {\bf 31}, 2664 (1990).
\bibitem{GMNO91} M. Gadella, M. A. Mart\'{\i}n, L. M. Nieto, and
        M. A. del Olmo, {\em ibid.} {\bf 32}, 1182 (1991).
\bibitem{BGO92} A. Ballesteros, M. Gadella, and M. A. Del Olmo,
        {\em ibid.} {\bf 33}, 3379 (1992).
\bibitem{MaOl96} M. A. Mart\'{\i}n and M. A. del Olmo, J. Phys. A 
        {\bf 29}, 689 (1996).
\bibitem{ArOl97a} O. Arratia and M. A. del Olmo, Fortschr. Phys. 
        {\bf 45}, 103 (1997).
\bibitem{ArOl97b} O. Arratia and M. A. del Olmo, Rep. Math. Phys. 
        {\bf 40}, 149 (1997).
\bibitem{GaNi} M. Gadella and L. M. Nieto, J. Phys. A {\bf 26},
        6043 (1993); 
        Fortschr. Phys. {\bf 42}, 261 (1994).
%
\bibitem{BM98} C. Brif and A. Mann, J. Phys. A {\bf 31}, L9 (1998).
%
\bibitem{Per} A. M. Perelomov, Commun. Math. Phys. {\bf 26}, 
        222 (1972); 
        {\em Generalized Coherent States and Their Applications\/}
        (Springer, Berlin, 1986).
%
\bibitem{Ber75} F. A. Berezin, Commun. Math. Phys. {\bf 40}, 
        153 (1975).
%
\bibitem{BMR98} C. Brif, A. Mann, and M. Revzen, 
        Phys. Rev. A {\bf 57}, 742 (1998).
%
\bibitem{Kiril} A. A. Kirillov, {\em Elements of the Theory of 
         Representations\/} (Springer, Berlin, 1976).
%
\bibitem{OCWi81} R. F. O'Connell and E. P. Wigner, Phys. Lett. A
        {\bf 83}, 145 (1981).
%
\bibitem{BaRa86} A. O. Barut and R. Raczka, {\em Theory of Group 
        Representations and Applications}, 2nd ed. 
        (World Scientific, Singapore, 1986), Chap. 15.
%
\bibitem{signofs} In Ref.\ \cite{BM98} we used a definition of
        the SW kernel differing from the present one by the 
        sign of $s$. In the present paper we choose the sign of 
        $s$ in accordance with the convention accepted in the
	quantum optics literature.
%
\bibitem{Agar81} G. S. Agarwal, Phys. Rev. A {\bf 24}, 2889 (1981).
%
\bibitem{note1} However, it should be noted that to the best of 
        our knowledge the whole $s$-parameterized family of 
	phase-space functions for SU(2) was not presented explicitly 
	in the literature before the work \cite{BM98}.
%
\bibitem{KaSu69} D. M. Kaplan and G. C. Summerfield, Phys. Rev.
        {\bf 187}, 639 (1969).
%
\bibitem{Gilm} R. Gilmore, in {\em Lecture Notes in Physics}, 
        vol. 278, edited by Y. S. Kim and W. W. Zachary, 
        (Springer, Berlin, 1987), p. 211; 
        W.-M. Zhang, D. H. Feng, and R. Gilmore,
        Rev. Mod. Phys. {\bf 62}, 867 (1990).
%
\bibitem{DAS94} J. P. Dowling, G. S. Agarwal, and W. P. Schleich,
        Phys. Rev. A {\bf 49}, 4101 (1994).
%
\bibitem{KBWolf} K. B. Wolf, Opt. Commun. {\bf 132}, 343 (1996).
%
\bibitem{CziBen} A. Czirj\'{a}k and M. G. Benedict, 
        Quantum Semiclass. Opt. {\bf 8}, 975 (1996);
        Acta Phys. Slov. {\bf 47}, 263 (1997).
%
\bibitem{Woot87} W. K. Wootters, Ann. Phys. (N.Y.) {\bf 176}, 
        1 (1987).
\bibitem{Leonh} U. Leonhardt, Phys. Rev. Lett. {\bf 74},
        4101 (1995); 
        Phys. Rev. A {\bf 53}, 2998 (1996).
%
\bibitem{Klaud} J. R. Klauder, J. Math. Phys. {\bf 4}, 1058 (1963). 
%
\bibitem{ACGT72} F. T. Arecchi, E. Courtens, R. Gilmore, and 
        H. Thomas, Phys. Rev. A {\bf 6}, 2211 (1972).
%
\bibitem{Fano} U. Fano, Phys. Rev. {\bf 90}, 577 (1953).
%
\bibitem{Fano57} U. Fano, Rev. Mod. Phys. {\bf 29}, 74 (1957).
\bibitem{Pauli58} W. Pauli, in {\em Encyclopedia of Physics\/}
        (Springer, Berlin, 1958), Vol. 5, p. 17.
\bibitem{Rec:early1} W. Gale, E. Guth, and G. T. Trammell,
        Phys. Rev. {\bf 165}, 1434 (1968);
        A. Vogt, in {\em Mathematical Foundations of Quantum 
        Theory}, edited by A. R. Marlow 
        (Academic, New York, 1978), p. 365.
        J. V. Corbett and C. A. Hurst, J. Austral. Math. Soc. B
        {\bf 20}, 182 (1978);
        A. Or{\l}owski and H. Paul, Phys. Rev. A {\bf 50}, 
        R921 (1994).
\bibitem{BaPa} W. Band and J. L. Park, Found. Phys. {\bf 1},
        133 (1970); {\bf 1}, 211 (1971); {\bf 1}, 339 (1971);
        Am. J. Phys. {\bf 47}, 188 (1979);
        I. D. Ivanovi\'{c}, J. Math. Phys. {\bf 24}, 1199 (1983).
\bibitem{Woot} W. K. Wootters, Found. Phys. {\bf 16}, 391 (1986);
        W. K. Wootters and B. D. Fields, Ann. Phys. (N.Y.)
        {\bf 191}, 363 (1989).
%
\bibitem{Prug77} E. Prugove\v{c}ki, Int. J. Theor. Phys. {\bf 16},
        321 (1977).
\bibitem{Busch91} P. Busch, Int. J. Theor. Phys. {\bf 30},
        1217 (1991).
\bibitem{HeSch95} D. M. Healy, Jr. and F. E. Schroeck, Jr.,
        J. Math. Phys. {\bf 36}, 453 (1995).
%
\bibitem{BeBe87} J. Bertrand and P. Bertrand, Found. Phys. {\bf 17},
        397 (1987).
\bibitem{VoRi89} K. Vogel and H. Risken, Phys. Rev. A {\bf 40},
        2847 (1989).
\bibitem{Raymer} D. T. Smithey, M. Beck, M. G. Raymer, and 
        A. Faridani, Phys. Rev. Lett. {\bf 70}, 1244 (1993);
        D. T. Smithey, M. Beck, J. Cooper, M. G. Raymer, and 
        A. Faridani, Phys. Scr. {\bf T48}, 35 (1993);
        M. Munroe, D. Boggavarapu, M. E. Anderson, and 
        M. G. Raymer, Phys. Rev. A {\bf 52}, R924 (1995).
\bibitem{Mlynek} S. Schiller, G. Breitenbach, S. F. Pereira,
        T. M\"{u}ller, and J. Mlynek, Phys. Rev. Lett. {\bf 77},
        2933 (1996). 
        G. Breitenbach and S. Schiller, J. Mod. Opt.
        {\bf 44}, 2207 (1997);
        G. Breitenbach, S. Schiller, and J. Mlynek,
        Nature (London) {\bf 387}, 471 (1997).
\bibitem{DAMP94} G. M. D'Ariano, C. Macchiavello, and 
        M. G. A. Paris, Phys. Lett A {\bf 195}, 31 (1994);
        Phys. Rev. A {\bf 50}, 4298 (1994).
\bibitem{KWV94} H. K\"{u}hn, D.-G. Welsh, and W. Vogel,
        J. Mod. Opt. {\bf 41}, 1607 (1994).
\bibitem{LPDA95} U. Leonhardt, H. Paul, and G. M. D'Ariano,
        Phys. Rev. A {\bf 52}, 4899 (1995).
\bibitem{LMKRR96} U. Leonhardt, M. Munroe, T. Kiss, Th. Richter,
        and M. G. Raymer, Opt. Commun. {\bf 127}, 144 (1996);
        U. Leonhardt and M. Munroe, 
        Phys. Rev. A {\bf 54}, 3682 (1996).
\bibitem{Wun97} A. W\"{u}nsche, J. Mod. Opt. {\bf 44}, 2293 (1997).
\bibitem{KWV95} H. K\"{u}hn, D.-G. Welsh, and W. Vogel,
        Phys. Rev. A {\bf 51}, 4240 (1995).
\bibitem{RMAL96} M. G. Raymer, D. F. McAlister, and U. Leonhardt,
        Phys. Rev. A {\bf 54}, 2397 (1996).
\bibitem{OWV97} T. Opatrn\'{y}, D.-G. Welsh, and W. Vogel,
        Opt. Commun. {\bf 134}, 112 (1997).
\bibitem{Richter97} Th. Richter, 
        J. Mod. Opt. {\bf 44}, 2385 (1997);
        Phys. Rev. A {\bf 55}, 4629 (1997).
\bibitem{PTKJ97} H. Paul, P. Torma, T. Kiss, and I. Jex,
        J. Mod. Opt. {\bf 44}, 2395 (1997).
%
\bibitem{symptom} S. Mancini, V. I. Man'ko, and P. Tombesi,
        Quantum Semiclass. Opt. {\bf 7}, 615 (1995);
        G. M. D'Ariano, S. Mancini, V. I. Man'ko, and P. Tombesi,
        {\em ibid.} {\bf 8}, 1017 (1996);
        V. I. Man'ko, J. Russ. Laser Research {\bf 17}, 579 (1996).
%
\bibitem{BaWo96} K. Banaszek and K. W\'{o}dkiewicz,
        Phys. Rev. Lett. {\bf 76}, 4344 (1996).
\bibitem{WaVo96} S. Wallentowitz and W. Vogel,
        Phys. Rev. A {\bf 53}, 4528 (1996).
\bibitem{OpWe97} T. Opatrn\'{y} and D.-G. Welsh, 
        Phys. Rev. A {\bf 55}, 1462 (1997);
        T. Opatrn\'{y}, D.-G. Welsh, S. Wallentowitz, 
        and W. Vogel, J. Mod. Opt. {\bf 44}, 2405 (1997).
\bibitem{MTM97} S. Mancini, P. Tombesi, and V. I. Man'ko, 
        Europhys. Lett. {\bf 37}, 79 (1997).
%
\bibitem{BMS95} P. J. Bardroff, E. Mayr, and W. P. Schleich,
        Phys. Rev. A {\bf 51}, 4963 (1995);
        P. J. Bardroff, E. Mayr, W. P. Schleich, P. Domokos,
        M. Brune, J. M. Raimond, and S. Haroche,
        {\em ibid.} {\bf 53}, 2736 (1996).
\bibitem{LuDa97} L. G. Lutterbach and L. Davidovich,
        Phys. Rev. Lett. {\bf 78}, 2547 (1997). 
\bibitem{BAKW98} C. T. Bodendorf, G. Antesberger, M. S. Kim,
        and H. Walther, Phys. Rev. A {\bf 57}, 1371 (1998).
%
\bibitem{WaVo} S. Wallentowitz and W. Vogel,
        Phys. Rev. Lett. {\bf 75}, 2932 (1995);
        Phys. Rev. A {\bf 54}, 3322 (1996).
\bibitem{PWCZB96} J. F. Poyatos, R. Walser, J. I. Cirac, 
        P. Zoller, and R. Blatt, 
        Phys. Rev. A {\bf 53}, R1966 (1996).
\bibitem{DHMi96} C. D'Helon and G. J. Milburn,
        Phys. Rev. A {\bf 54}, R25 (1996)
\bibitem{BLSS96} P. J. Bardroff, C. Leichtele, G. Schrade,
        and W. P. Schleich,
        Phys. Rev. Lett. {\bf 77}, 2198 (1996).
\bibitem{Frey97} M. Freyberger, Phys. Rev. A {\bf 55},
        4120 (1997).
\bibitem{Leibfr} D. Leibfried, D. M. Meekhof, B. E. King,
        C. Monroe, W. M. Itano, and D. J. Wineland,
        Phys. Rev. Lett. {\bf 77}, 4281 (1996);
        D. Leibfried, D. M. Meekhof, C. Monroe, B. E. King,
        W. M. Itano, and D. J. Wineland,
        J. Mod. Opt. {\bf 44}, 2485 (1997).
%
\bibitem{Royer} A. Royer, Phys. Rev. Lett. {\bf 55}, 2745 (1985);
        Found. Phys. {\bf 19}, 3 (1989).
\bibitem{1dwp} U. Leonhardt and M. G. Raymer, 
        Phys. Rev. Lett. {\bf 76}, 1985 (1996);
        D. S. Kraehmer and U. Leonhardt, 
        Appl. Phys. B {\bf 65}, 725 (1997);
        U. Leonhardt and S.~Schneider, 
        Phys. Rev. A {\bf 56}, 2549 (1997);
        U. Leonhardt and P. J. Bardroff,
        Acta Phys. Slov. {\bf 47}, 225 (1997);
        M.~G.~Raymer, J. Mod. Opt. {\bf 44}, 2565 (1997).
\bibitem{anharm} U. Leonhardt, Phys. Rev. A {\bf 55}, 3164 (1997);
        T. Opatrn\'{y}, D.-G. Welsh, and W. Vogel,
        {\em ibid.} {\bf 56}, 1788 (1997).
\bibitem{DOZ98} L. Davidovich, M. Orszag, and N. Zagury,
        Phys. Rev. A {\bf 57}, 2544 (1998).
\bibitem{atbeams} U. Janicke and M. Wilkens,
        J. Mod. Opt. {\bf 42}, 2183 (1995);
	Ch. Kurtsiefer, T. Pfau, and J. Mlynek,
        Nature (London) {\bf 386}, 150 (1997);
	T. Pfau and Ch. Kurtsiefer, J. Mod. Opt.
        {\bf 44}, 2551 (1997);
        D. A. Kokorowski and D. E. Pritchard, {\em ibid.}
        {\bf 44}, 2575 (1997);
        S. H. Kienle, D. Fischer, W. P. Schleich, V. P. Yakovlev,
        and M. Freyberger, Appl. Phys. B {\bf 65}, 735 (1997).
%
\bibitem{beccond} S. Mancini and P. Tombesi,
        Europhys. Lett. {\bf 40}, 351 (1997);
        E. L. Bolda, S. M. Tan, and D. F. Walls,
        Phys. Rev. Lett. {\bf 79}, 4719 (1997);
	Phys. Rev. A {\bf 57}, 4686 (1998).
%
\bibitem{MaTo97} S. Mancini and P. Tombesi, 
        Phys. Rev. A {\bf 56}, 3060 (1997).
\bibitem{ChYe97} X. Chen and J. A. Yeazell,
        Phys. Rev. A {\bf 56}, 2316 (1997).
%
\bibitem{DoMa97} V. V. Dodonov and V. I. Man'ko,
        Phys. Lett. A {\bf 229}, 335 (1997).
\bibitem{Agar98} G. S. Agarwal, 
        Phys. Rev. A {\bf 57}, 671 (1998).
%
\bibitem{ACBWR90} J. R. Ashburn, R. A. Cline, 
        P. J. M. van der Burgt, W. B. Westerveldt, and
        J. S. Risley, Phys. Rev. A {\bf 41}, 2407 (1990).
\bibitem{molvibr} T. J. Dunn, J. N. Sweetser, I. A. Walmsley,
        and C. Radzewicz, 
        Phys. Rev. Lett. {\bf 70}, 3388 (1993);
        T. J. Dunn, I. A. Walmsley, and S. Mukamel, 
        {\em ibid.} {\bf 74}, 884 (1995).
%
\bibitem{Wod84} K. W\'{o}dkiewicz, Phys. Rev. Lett. {\bf 52},
	1064 (1984);
	Phys. Lett. A {\bf 115}, 304 (1986).
\bibitem{BKK95} V. Bu\v{z}ek, C. H. Keital, and P. L. Knight,
	Phys. Rev. A {\bf 51}, 2575 (1995);
	{\bf 51}, 2594 (1995).
\bibitem{Ban98} M. Ban, Int. J. Theor. Phys. {\bf 38}, 
	2583 (1998);
	J. Math. Phys. {\bf 39}, 1744 (1998).
%
\bibitem{MCKn93} H. Moya-Cessa and P. L. Knight,
        Phys. Rev. A {\bf 48}, 2479 (1993).
%
\bibitem{Wehrl} A. Wehrl, Rev. Mod. Phys. {\bf 50}, 221 (1978).
%
\bibitem{Trees68} H. L. van Trees, {\em Detection, Estimation,
        and Modulation Theory}, Vol. III 
        (Wiley, New York, 1968).
%

\end{references}
\end{document}